\documentclass[aps,pre,showpacs,reprint,superscriptaddress,amsmath,amsfonts]{revtex4-1}
\usepackage{amssymb}
\usepackage{graphicx}
\allowdisplaybreaks
\usepackage{pgfplots}
\usepackage{stackrel}
\usepackage{color}

\definecolor{mygreen}{RGB}{50,205,50}
\definecolor{myyellow}{RGB}{255,215,50}

\newcommand*\kB{\ensuremath{k_\mathrm{B}}}

\renewcommand{\tensor}[1]{\boldsymbol{#1}}
\renewcommand{\vec}[1]{\boldsymbol{#1}}
\newcommand{\diff}{\ensuremath{\mathrm d}}

\newcommand*\thbm{\ensuremath{T_{\mathrm{HBM}}}}

\usepackage{mathrsfs}

\begin{document}

\title{Effective temperatures of hot Brownian motion} 

\newcommand\mpi{\affiliation{Max Planck Institute for Mathematics in the Sciences, Inselstr. 22,
04103 Leipzig, Germany}}
\newcommand\ul{\affiliation{Institut f\"ur Theoretische Physik, Universit\"at Leipzig,  Postfach 100 920, D-04009 Leipzig, Germany}}
\newcommand\ule{\affiliation{Department of Physics and Astronomy, University of Leeds, LS2 9JT Leeds, England}}
\author{G.~Falasco}\ul
\author{M.~V.~Gnann}\mpi
\author{D.~Rings}\ule
\author{K.~Kroy}\ul

\begin{abstract}
  We derive generalized Langevin equations for the translational and
  rotational motion of a heated Brownian particle from the fluctuating
  hydrodynamics of its non-isothermal solvent. The temperature
  gradient around the particle couples to the hydrodynamic modes
  excited by the particle itself so that the resulting noise spectrum
  is governed by a frequency-dependent temperature.  We show how the
  effective temperatures at which the particle coordinates and
  (angular) velocities appear to be thermalized emerge from this
  central quantity.
\end{abstract}

\maketitle

\section{Introduction}
Hot Brownian motion \cite{Rings.2010} is the diffusive dynamics of a
colloidal particle persistently maintained at higher temperature than
the surrounding fluid, so that the fluid temperature field $T(\vec
r)=T(r)$ decays radially around the particle. It is of practical
relevance, e.g. for laser-heated suspended nanoparticles involved in
several experimental applications ranging from particle trapping and
tracking \cite{Berciaud.2004, Braun.2013} to self-thermophoretic micro-swimmers \cite{Sano.2010,
  Qian.2013}. Besides, it is also of considerable theoretical
interest, since it can be thought of as an archetypical example of a
system in contact with a non-isothermal bath, hence far from thermal
equilibrium. Nevertheless, for important conceptual and practical
purposes, the hot particle can often be treated like an equivalent
Brownian particle in equilibrium, with appropriate effective transport
coefficients.

In particular, it has been shown analytically \cite{Chakraborty.2011,
  Rings.2012} that free and confined diffusion of a hot spherical
particle are, in the long-time limit, governed by effective
``positional'' temperatures, denoted by $T^{\scriptscriptstyle X}$ and $T^{\scriptscriptstyle \Theta}$ for
translation in the $X-$direction and rotation along an angle $\Theta$,
respectively. The positional temperatures enter the effective
Stokes--Einstein relations and Boltzmann factors for translation and
rotation of the particle, respectively.  Additionally, extensive
numerical simulations \cite{Barrat.2011, Chakraborty.2011} have shown
that the Maxwellian (angular) velocity distribution and the short-time
response of the hot Brownian particle are characterized by yet other,
somewhat higher effective temperatures, so-called kinetic temperatures
$T^{\scriptscriptstyle V}$ and $T^{\scriptscriptstyle \Omega}$. None of these effective temperatures is
generally equal to the solvent temperature $T_s\equiv  T(r\to R)$ at the
particle surface or to the ambient temperature $T_0\equiv  T(r\to
\infty)$.  This complex behavior has led to the conclusion that an
effective Langevin description of hot Brownian motion is restricted to
the Markov limit \cite{Barrat.2011}.

Here, we show constructively how this limitation can be overcome,
starting from the fluctuating hydrodynamics of a solvent maintained at
local thermal equilibrium with a temperature field $T(\vec r)$. On
this basis, we derive the generalized Langevin equation (GLE) for a
heated spherical particle.  Conceptually, non-spherical particles can
be treated along the same lines, albeit with additional complications
\cite{Rings.2012}.  The most conspicuous feature of the theory is a
frequency-dependent noise temperature $\mathcal{T}(\omega)$ \cite{Falasco.2014}. It arises from the hydrodynamic
coupling between the particle and distant solvent volume elements that
are locally equilibrated at different temperatures $T(\vec r)$. From this central quantity analytical predictions for the mentioned kinetic and positional
effective temperatures are derived.

The characteristic frequency scales that primarily select the dominant
modes from the ``temperature spectrum'' $\mathcal{T}(\omega)$ are (for
a translating sphere of mass $M$, density $\varrho_\mathrm{p}$, and
radius $R$)
\begin{equation}\label{eq:rates}
  \omega_{\text{f}} \equiv  \frac{2\nu}{R^2} \quad \text{and} \quad  \omega_\mathrm{p} \equiv  
  \frac{6\pi \eta R}{M} = \frac{9\varrho}{4\varrho_\mathrm{p}}\omega_\mathrm{f}\;,
\end{equation}
namely, the inverse time scale for vorticity diffusion across the
particle, and the inverse Stokes relaxation time of the particle
momentum, respectively. The former characterizes how efficiently the
particle momentum is spatially dispersed in a solvent of kinematic
viscosity $\nu=\eta/\varrho$ and density $\varrho$, and the latter
how, as a result, the motion of the particle adjusts to that of the
fluid. The meaning of slow and fast processes, or low and high
frequencies of the noise spectrum, is primarily provided by these
rates.  It should be clear, though, that any externally imposed
additional time scale that interferes with these rates, can be
expected to yield additional features.

The paper is structured as follows. In the next section we introduce
the theoretical model of a Brownian particle in a non-isothermal
solvent. We then sketch the contraction of the coupled
solvent-particle system to the GLE for the particle motion, alone.
Details of the calculation are given in Appendix
\ref{sec:derivation}. In Sec.~\ref{sec:Tomega}, we examine the
frequency-dependent temperatures $\mathcal{T}(\omega)$ that govern the
Langevin noise for the translational and rotational degrees of freedom
of a heated sphere and give a qualitative physical interpretation of
their functional form, while some technicalities are deferred to
Appendix \ref{sec:hydro}. From this central quantity, we derive the
effective rotational and translational kinetic temperatures of a free
particle in Sec.~\ref{sec:Tkin}. We analyze their explicit dependence
on the characteristic time scales for the velocity relaxation of the
particle and the solvent by varying their density ratio. Also we regain the known positional temperatures for
translation and rotation \cite{Chakraborty.2011,Rings.2012} as the low
frequency limits of $\mathcal{T}(\omega)$. In Section
\ref{sec:pos}, we consider a hot Brownian particle trapped in a
harmonic potential. While the known effective-equilibrium description
can be retrieved in the Markov limit, we point out that the kinetic
and positional temperatures may differ from those in the free case for
very stiff traps, due to the interference of the characteristic trap
time scale with the rates defined in Eq.~\eqref{eq:rates}.  We conclude
with a summary and short outlook.

\section{Fluctuating hydrodynamics of a heated particle}\label{sec:fh}

We consider a spherical particle of radius $R$ immersed in an
incompressible fluid of density $\varrho$ described by the linearized
fluctuating hydrodynamic equations \cite{Landau.1987, Hauge.1973}
\begin{subequations}\label{sys}
\begin{eqnarray}
&&\varrho \partial_t \vec v(\vec r, t) - \nabla \cdot \tensor \sigma (\vec r, t) = \nabla \cdot \tensor \tau (\vec r, t)\,,\label{sys1}\\
&&\nabla \cdot \vec v (\vec r, t) = 0\,, \label{sys2}\\
&&\vec v(\vec r, t)= \vec V(t) + \vec \Omega(t) \times \vec r\, \text{ on } \mathcal{S} , \label{sys3}
\end{eqnarray}
\end{subequations}
where the velocity field $\vec v$ of the fluid is defined in the
volume $\mathcal{V}$ outside the particle and the no-slip boundary condition on
the particle surface $\mathcal{S}$ is imposed by Eq.~\eqref{sys3}. The stress
tensor $\vec \sigma$ has components $\sigma_{ij}= -p \delta_{ij} +2
\eta \Gamma_{ij}$, where $p$ is the pressure and $\Gamma_{ij}=
(\partial_i v_j + \partial_j v_i)/2$ the shear rate tensor, with the
dynamic viscosity $\eta$. The incompressibility condition Eq.~\eqref{sys2}
can be eliminated by expressing $p$ (and thus $\vec \sigma$) as a
functional of the flow field $\vec v$. Finally, the thermal noise is
represented by a zero-mean Gaussian random stress tensor
$\tensor{\tau}$ that vanishes on the particle surface and otherwise
obeys the fluctuation-dissipation relation
\begin{eqnarray}\nonumber
\left< \tau_{ij}(\vec r, t)  \tau_{kl}(\vec r^\prime, t^\prime) \right> &=& 2 \eta(\vec r, t) \kB T(\vec r, t) \delta(\vec r - \vec r^\prime)\\
&& \times \delta (t - t^\prime) \left(\delta_{ik} \delta_{jl} + \delta_{il} \delta_{jk} \right) \label{fluidfdt}
\end{eqnarray}
corresponding to a local equilibrium with the prescribed temperature
field $T(\vec r,t)$ \cite{Sengers.2006}.

The vectors $\vec V(t)$ and $\vec \Omega(t)$, denoting the
translational and rotational velocity of the Brownian particle, couple
to the solvent dynamics via the boundary condition Eq.~\eqref{sys3} on the
particle surface $\mathcal{S}$. They evolve themselves according to Newton's
equations of motion
\begin{subequations}\label{newton}
\begin{eqnarray}
M \dot{\vec V}(t) &=& \vec F(t) +  \vec{F}_{\text{e}}(t)\,, \label{newton1}\\
I \dot{\vec \Omega}(t) &=& \vec T(t) +\vec{T}_{\text{e}}(t)\,, \label{newton2}
\end{eqnarray}
\end{subequations}
where $M$ is the mass of the particle, $I$ the moment of inertia,
$\vec{F}_{\text{e}}$ and $\vec{T}_{\text{e}}$ are the external
force and torque, and $\vec{F}$ and $\vec{T}$ are the hydrodynamic
force and torque exerted by the fluid, defined by
\begin{subequations}\label{formom}
\begin{eqnarray}
\vec F(t) &=&- \int_{\mathcal{S}} \tensor \sigma (\vec r, t) \cdot \vec n \, \diff^2 r\,,\\
\vec T(t) &=& - \int_{\mathcal{S}} \vec r \times \left(\tensor \sigma (\vec r, t) \cdot \vec n\right) \diff^2 r\,,
\end{eqnarray}
\end{subequations}
with $\vec{n}$ the inner radial unit vector.  Note that we have
suppressed the time dependence of $\mathcal{S}$ in Eqs.~\eqref{sys3},~\eqref{formom} in order to make the above set of equations linear not
only in the flow field but also in the particle velocity. See \cite{Hinch.1993, Hermans.1981} for a discussion of the validity of linear hydrodynamics in relation to Brownian motion. We also suppress the corresponding time-dependent
thermal advection, by requiring the temperature field to obey the
stationary heat equation in the co-moving frame,
\begin{eqnarray}\label{heateq}
&&\nabla^2 T(\vec r)=0\,,\\
&&T(\vec r)= T_0+\Delta T \text{ on } \mathcal{S}\,, \nonumber\\
&&T(r \to \infty)= T_0\,. \nonumber
\end{eqnarray}
This technical simplification and other implicit idealizations, such
as taking the heat conductivity of the solvent to be constant, can be justified for common
experimental conditions, such as those realized for laser-heated
nanoparticles in water \cite{Rings.2010, Rings.2011}. Together with
the prescription Eq.~\eqref{heateq}, the system
Eq.~\eqref{sys}--Eq.~\eqref{fluidfdt} then entirely describes the time
evolution of the fluid and the heated Brownian particle. The solution
of Eq.~\eqref{heateq} is the radial field:
\begin{equation}\label{eq:Tfourier}
T(r)= T_0+\Delta T R/r\,. 
\end{equation}
While the following derivation does not
strictly depend on the specific form of $T(\vec r)$ (as long as it
does not depend on the particle velocity), and even an explicit
externally imposed dependence on time could be included, we restrict
the discussion in the following sections to this paradigmatic case.

We now proceed to contract the description of fluid plus particle into
an equation for the particle alone. We rewrite the hydrodynamic forces
introduced in Eq.~\eqref{newton} in the form
\begin{subequations}
\begin{eqnarray}
\vec F &\equiv  \vec F_{\text{d}}+ \vec \xi^{\text{\tiny{T}}}  \\
\vec T &\equiv  \vec T_{\text{d}}+ \vec \xi^{\text{\tiny{R}}}
\end{eqnarray}
\end{subequations}
to account for contributions $\vec \xi$ independent of the particle
velocity that are expected to arise due to the inhomogeneity of
Eq.~\eqref{sys1}. By Eq.~\eqref{sys}, $\vec v(\vec r, t)$ is a linear
functional of $\vec V(t^\prime)$ and $\vec \Omega(t^\prime)$ with
$-\infty <t^\prime <t$, so in view of Eq.~\eqref{formom} this implies
that the systematic components $\vec F_{\text{d}}$ and $\vec
T_{\text{d}}$ are linear functionals of $\vec V(t^\prime)$ and $\vec
\Omega(t^\prime)$ respectively, with $-\infty <t^\prime <t$. Hence, we
can write
\begin{eqnarray}
\vec F_{\text{d}}(t) &=& - \int_{-\infty}^t \zeta(t - t^\prime)  \vec V(t^\prime) \diff t^\prime,\label{sysforce1}\\
\vec T_{\text{d}}(t) &=& - \int_{-\infty}^t \gamma(t - t^\prime) \vec \Omega(t^\prime) \diff t^\prime,\label{sysforce2}
\end{eqnarray}
where $\zeta(t)$ and $\gamma(t)$ are positive, time-symmetric memory
kernels accounting for the time-dependent drag on the particle
\cite{Hauge.1973}. Equations \eqref{newton} then take the GLE form
\begin{eqnarray}
&&M \dot{\vec V}(t) =  - \!\int_{-\infty}^t\!\!\!\! \zeta(t - t^\prime)  \vec V(t^\prime) \diff t^\prime + \vec \xi^{\text{\tiny{T}}}(t)+  \vec{F}_{\text{e}}(t), \label{GLEt}\\
&&I \dot{\vec \Omega}(t) =  - \!\int_{-\infty}^t\!\!\!\! \gamma(t - t^\prime) \vec \Omega(t^\prime) \diff t^\prime +\vec \xi^{\text{\tiny{R}}}(t)+ \vec{T}_{\text{e}}(t), \,\,\,\label{GLEr}
\end{eqnarray}
once we identify $\vec \xi^{\text{\tiny{T}},\text{\tiny{R}}}$ as the
Langevin noise, whose statistical properties have to be derived from
those of the random stress tensor $\tensor \tau$. 

For better readability, the actual calculation is detailed in
Appendix~\ref{sec:derivation}, and only the main results and their
physical interpretations are given in the main text.  We focus mostly
on the translational motion, but the rotational case is very
analogous. It is moreover convenient to switch to the frequency
representation defining, for a generic function $g(\omega)$, the
Fourier transform $g(\omega) \equiv  \int_{-\infty}^\infty e^{i \omega t}
g(t) \diff t$ and the half-Fourier transform
$g^\text{\tiny{+}}(\omega) \equiv  \int_{0}^\infty e^{i \omega t} g(t)
\diff t$. 

To complete the contraction, we compare the energy dissipated by the
fluid friction acting on the particle at a mean velocity $\langle \vec
V(\omega) \rangle$
\begin{equation}\label{result1}
\zeta(\omega)\delta_{ij}\langle V_i(\omega)\rangle \langle V_j^*(\omega)\rangle=2 \int_{\mathcal{V}} \phi^{\text{\tiny{T}}}(\vec r, \omega) \, \diff^3 r\,,
\end{equation}
with the correlation function of the energy supplied by the random
force at frequencies $\omega$ and $\omega^\prime$
\begin{eqnarray}\nonumber
&&\langle \xi_i^{\text{\tiny{T}}}(\omega) \xi_j^{\text{\tiny{T}}*}(\omega')\rangle \langle V_i(\omega)\rangle \langle V_j^*(\omega')\rangle =\\
&&= 2 \kB \delta(\omega-\omega') \int_{\mathcal{V}}  \phi^{\text{\tiny{T}}}(\vec
r, \omega) T(\vec r) \, \diff^3 r \,. \label{result2}
\end{eqnarray}
From Appendix~\ref{sec:derivation}, we have quoted the representation
in terms of the dissipation function,
\begin{equation}\label{diss}
  \phi^{\text{\tiny{T}}}(\vec r,
  \omega) \equiv   \eta \left(\partial_i u_j \partial_i u_j^*+ \partial_i
    u_j\partial_j u_i^*\right), 
\end{equation}
which gives the energy dissipated by the fluid at position $\vec r$ and
frequency $\omega$ in terms of the flow field $\vec u(\vec r,\omega)$.

From Eq.~\eqref{result1} and Eq.~\eqref{result2} we then find the relation \begin{eqnarray}\nonumber
&&\langle \xi_i^{\text{\tiny{T}}}(\omega)   \xi_j^{\text{\tiny{T}}*}(\omega')\rangle \langle V_i(\omega)\rangle \langle V_j^*(\omega')\rangle =\\ \label{correlation}
&&= \kB \mathcal{T}^{\text{\tiny{T}}}(\omega) \zeta(\omega) \delta_{ij}\delta(\omega-\omega') \langle V_i(\omega)\rangle \langle V_j^*(\omega')\rangle
\end{eqnarray}
with
\begin{equation}\label{Tdef}
\mathcal{T}^{\text{\tiny{T}}}(\omega)\equiv \frac{\int_{\mathcal{V}}  \phi^{\text{\tiny{T}}}(\vec r, \omega) T(\vec r) \, \diff^3 r}{\int_{\mathcal{V}}  \phi^{\text{\tiny{T}}}(\vec r, \omega)\, \diff^3 r}\,.
\end{equation}
Since $\phi^{\text{\tiny{T}}}(\vec r,\omega)$
is a quadratic function of $\langle \vec V(\omega)\rangle$ (see
Appendix \ref{sec:hydro}) the ratio in Eq.~\eqref{Tdef} is independent of
$\langle \vec V(\omega)\rangle$. Moreover, as the particle velocity
$\langle \vec V(\omega) \rangle$ is arbitrary it can be deleted in
Eq.~\eqref{correlation}, which renders Eq.~\eqref{correlation} in
the form of a generalized fluctuation-dissipation relation:
\begin{equation}\label{fdt}
\langle \xi_i^{\text{\tiny{T}}}(\omega)  \xi_j^{\text{\tiny{T}}*}(\omega')\rangle= \kB \mathcal{T}^{\text{\tiny{T}}}(\omega) \zeta(\omega)  \delta_{ij}\delta(\omega-\omega')\,.
\end{equation}
Analogous results hold for the rotational motion. They are obtained by
substituting $\zeta \to \gamma$ in Eq.~\eqref{fdt} and
$\phi^{\text{\tiny{T}}} \to \phi^{\text{\tiny{R}}}$ in the definition
Eq.~\eqref{Tdef}.

\section{The noise temperature $\mathcal{T}(\omega)$}\label{sec:Tomega}

Eq.~\eqref{Tdef} defines the frequency-dependent noise temperature that is the central quantity for the Brownian motion under non-isothermal conditions. Its nonlocal nature is manifest in the
weighted average over the temperature field $T(\vec r)$, with the
dissipation function determining how strongly the diverse local
temperatures in the surroundings affect the Brownian motion of the
particle at the origin.

Clearly, the noise autocorrelation can always be cast in such a form
by defining a suitable function $\mathcal{T}(\omega)$ that measures
the violation of the equilibrium fluctuation-dissipation
relation. Here, the nontrivial statement is that $\mathcal{T}(\omega)$
is explicitly derived from an underlying hydrodynamic
description. Moreover, in the next sections, we will show that
$\mathcal{T}(\omega)$ plays the role of a frequency-dependent
effective temperature, in the sense that dynamical isothermal
relations can directly be extended to the non-isothermal case if the
temperature $T_0$ is replaced by $\mathcal{T}(\omega)$.

\begin{figure}
\begin{tikzpicture}
\begin{semilogxaxis}[xmin=0.01,xmax=100,legend style={ at={(0.05,0.85)}, anchor=west}, xlabel=$\sqrt{\omega/ \omega_\text{f}}$,
ylabel=$\left(\mathcal{T}(\omega)-T_0\right)/\Delta T$] 
\addplot[color=red,mark=*, mark size = 0] coordinates {(0.001,0.417128)(0.011,0.421213)(0.021,0.425282)(0.031,0.429292)(0.041,0.433244)(0.051,0.437138)(0.061,0.440977)(0.071,0.444762)(0.081,0.448492)(0.091,0.452171)(0.101,0.455798)(0.111,0.459375)(0.121,0.462902)(0.131,0.466382)(0.141,0.469814)(0.151,0.4732)(0.161,0.476541)(0.171,0.479838)(0.181,0.483091)(0.191,0.486301)(0.201,0.48947)(0.211,0.492598)(0.221,0.495685)(0.231,0.498734)(0.241,0.501743)(0.251,0.504716)(0.261,0.50765)(0.271,0.510549)(0.281,0.513412)(0.291,0.51624)(0.301,0.519034)(0.311,0.521794)(0.321,0.524521)(0.331,0.527215)(0.341,0.529878)(0.351,0.532509)(0.361,0.53511)(0.371,0.53768)(0.381,0.540221)(0.391,0.542733)(0.401,0.545216)(0.411,0.547671)(0.421,0.550099)(0.431,0.5525)(0.441,0.554874)(0.451,0.557222)(0.461,0.559544)(0.471,0.561841)(0.481,0.564113)(0.491,0.56636)(0.501,0.568584)(0.511,0.570784)(0.521,0.57296)(0.531,0.575114)(0.541,0.577245)(0.551,0.579355)(0.561,0.581442)(0.571,0.583508)(0.581,0.585553)(0.591,0.587577)(0.601,0.589581)(0.611,0.591565)(0.621,0.593529)(0.631,0.595473)(0.641,0.597399)(0.651,0.599305)(0.661,0.601193)(0.671,0.603062)(0.681,0.604914)(0.691,0.606747)(0.701,0.608564)(0.711,0.610362)(0.721,0.612144)(0.731,0.61391)(0.741,0.615658)(0.751,0.617391)(0.761,0.619107)(0.771,0.620808)(0.781,0.622493)(0.791,0.624162)(0.801,0.625817)(0.811,0.627456)(0.821,0.629081)(0.831,0.630691)(0.841,0.632288)(0.851,0.633869)(0.861,0.635437)(0.871,0.636992)(0.881,0.638532)(0.891,0.64006)(0.901,0.641574)(0.911,0.643075)(0.921,0.644563)(0.931,0.646039)(0.941,0.647502)(0.951,0.648953)(0.961,0.650392)(0.971,0.651818)(0.981,0.653233)(0.991,0.654636)(1,0.655889)(2,0.754371)(3,0.808934)(4,0.843731)(5,0.867875)(6,0.885605)(7,0.899173)(8,0.909887)(9,0.918559)(10,0.925719)(11,0.931731)(12,0.936848)(13,0.941256)(14,0.945092)(15,0.94846)(16,0.951441)(17,0.954098)(18,0.95648)(19,0.958628)(20,0.960575)(21,0.962347)(22,0.963968)(23,0.965455)(24,0.966825)(25,0.96809)(26,0.969263)(27,0.970353)(28,0.971368)(29,0.972317)(30,0.973204)(31,0.974037)(32,0.974819)(33,0.975556)(34,0.976251)(35,0.976908)(36,0.977529)(37,0.978118)(38,0.978677)(39,0.979208)(40,0.979713)(41,0.980195)(42,0.980654)(43,0.981092)(44,0.981511)(45,0.981912)(46,0.982296)(47,0.982663)(48,0.983016)(49,0.983355)(50,0.983681)(51,0.983994)(52,0.984295)(53,0.984585)(54,0.984865)(55,0.985135)(56,0.985395)(57,0.985646)(58,0.985889)(59,0.986124)(60,0.986351)(61,0.986571)(62,0.986784)(63,0.98699)(64,0.98719)(65,0.987383)(66,0.987571)(67,0.987754)(68,0.987931)(69,0.988103)(70,0.988271)(71,0.988433)(72,0.988592)(73,0.988746)(74,0.988896)(75,0.989041)(76,0.989184)(77,0.989322)(78,0.989457)(79,0.989589)(80,0.989717)(81,0.989843)(82,0.989965)(83,0.990084)(84,0.990201)(85,0.990315)(86,0.990426)(87,0.990534)(88,0.990641)(89,0.990745)(90,0.990846)(91,0.990946)(92,0.991043)(93,0.991138)(94,0.991231)(95,0.991323)(96,0.991412)(97,0.9915)(98,0.991585)(99,0.99167)(100,0.991752)
};
\addplot[color=blue,mark=-, mark size = 0] coordinates {(0.001,0.75)(0.011,0.750001)(0.021,0.750004)(0.031,0.750013)(0.041,0.750028)(0.051,0.750052)(0.061,0.750085)(0.071,0.75013)(0.081,0.750186)(0.091,0.750255)(0.101,0.750337)(0.111,0.750433)(0.121,0.750543)(0.131,0.750668)(0.141,0.750808)(0.151,0.750962)(0.161,0.751132)(0.171,0.751318)(0.181,0.751518)(0.191,0.751734)(0.201,0.751964)(0.211,0.75221)(0.221,0.75247)(0.231,0.752745)(0.241,0.753035)(0.251,0.753338)(0.261,0.753656)(0.271,0.753987)(0.281,0.754332)(0.291,0.754689)(0.301,0.755059)(0.311,0.755442)(0.321,0.755837)(0.331,0.756243)(0.341,0.756661)(0.351,0.75709)(0.361,0.75753)(0.371,0.75798)(0.381,0.758441)(0.391,0.758911)(0.401,0.75939)(0.411,0.759878)(0.421,0.760376)(0.431,0.760881)(0.441,0.761395)(0.451,0.761917)(0.461,0.762446)(0.471,0.762982)(0.481,0.763525)(0.491,0.764075)(0.501,0.764631)(0.511,0.765193)(0.521,0.765761)(0.531,0.766334)(0.541,0.766912)(0.551,0.767496)(0.561,0.768084)(0.571,0.768676)(0.581,0.769273)(0.591,0.769874)(0.601,0.770478)(0.611,0.771086)(0.621,0.771697)(0.631,0.772312)(0.641,0.772929)(0.651,0.773549)(0.661,0.774171)(0.671,0.774796)(0.681,0.775423)(0.691,0.776052)(0.701,0.776683)(0.711,0.777315)(0.721,0.777949)(0.731,0.778584)(0.741,0.779221)(0.751,0.779858)(0.761,0.780496)(0.771,0.781135)(0.781,0.781775)(0.791,0.782415)(0.801,0.783055)(0.811,0.783696)(0.821,0.784337)(0.831,0.784978)(0.841,0.785619)(0.851,0.786259)(0.861,0.7869)(0.871,0.78754)(0.881,0.788179)(0.891,0.788818)(0.901,0.789457)(0.911,0.790094)(0.921,0.790731)(0.931,0.791367)(0.941,0.792002)(0.951,0.792636)(0.961,0.793269)(0.971,0.793901)(0.981,0.794532)(0.991,0.795161)(1,0.795726)(2,0.848126)(3,0.881524)(4,0.903438)(5,0.918697)(6,0.92987)(7,0.938381)(8,0.945069)(9,0.950459)(10,0.954893)(11,0.958602)(12,0.961751)(13,0.964457)(14,0.966806)(15,0.968865)(16,0.970685)(17,0.972304)(18,0.973754)(19,0.97506)(20,0.976242)(21,0.977318)(22,0.9783)(23,0.979201)(24,0.980031)(25,0.980797)(26,0.981506)(27,0.982165)(28,0.982778)(29,0.983351)(30,0.983887)(31,0.98439)(32,0.984862)(33,0.985306)(34,0.985726)(35,0.986122)(36,0.986496)(37,0.986851)(38,0.987188)(39,0.987508)(40,0.987812)(41,0.988102)(42,0.988378)(43,0.988642)(44,0.988894)(45,0.989135)(46,0.989366)(47,0.989588)(48,0.9898)(49,0.990004)(50,0.9902)(51,0.990388)(52,0.990569)(53,0.990744)(54,0.990912)(55,0.991074)(56,0.991231)(57,0.991382)(58,0.991528)(59,0.991669)(60,0.991805)(61,0.991938)(62,0.992065)(63,0.992189)(64,0.992309)(65,0.992426)(66,0.992539)(67,0.992649)(68,0.992755)(69,0.992859)(70,0.992959)(71,0.993057)(72,0.993152)(73,0.993244)(74,0.993335)(75,0.993422)(76,0.993508)(77,0.993591)(78,0.993672)(79,0.993751)(80,0.993828)(81,0.993903)(82,0.993977)(83,0.994048)(84,0.994118)(85,0.994187)(86,0.994254)(87,0.994319)(88,0.994383)(89,0.994445)(90,0.994506)(91,0.994566)(92,0.994624)(93,0.994681)(94,0.994737)(95,0.994792)(96,0.994846)(97,0.994898)(98,0.99495)(99,0.995)(100,0.99505)};
\draw ({axis cs:30,0}|-{rel axis cs:0,0}) -- ({axis cs:30,0}|-{rel axis cs:0,1});
\legend{$\mathcal{T}^{\text{\tiny{T}}}(\omega)$, $\mathcal{T}^{\text{\tiny{R}}}(\omega)$}
\end{semilogxaxis}
\end{tikzpicture}
\caption{The universal frequency-dependent noise temperatures for the motion of a Brownian particle, obtained by the definitions Eq.~\eqref{diss},~\eqref{Tdef} assuming a temperature-independent solvent  viscosity $\eta$.  The rotational noise temperature (\textcolor{blue}{---}) is given by the exact expression Eq.~\eqref{deltaT}, while the translational one (\textcolor{red}{---}) is obtained by numerical integration of  Eqs.~\eqref{Tdef},~\eqref{phi_transl}. The vertical solid line indicates the characteristic frequency beyond which the finite compressibility of water would matter for a solid particle of radius $R\simeq100\, \text{nm}$. }
\label{fig:Tomega}
\end{figure}
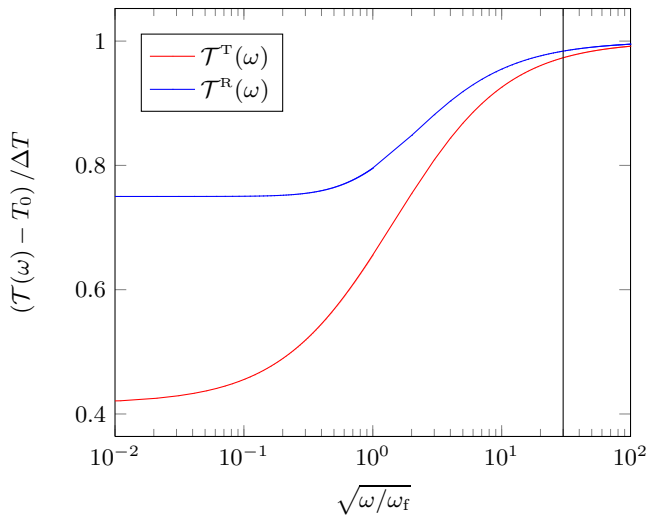

Contenting ourselves with explicit evaluations to leading order in the
temperature heterogeneity $T(\vec r)-T_0$, we can in the following
neglect a possible temperature-dependence of the viscosity, which
would affect our results to sub-leading order, only. Figure
\ref{fig:Tomega} shows the frequency-dependent temperatures
$\mathcal{T}(\omega)$ for the translational and the rotational motion
of a sphere, which are derived in Appendix~\ref{sec:hydro} assuming
constant heat conductivity and viscosity, i.e.\
Eq.~\eqref{heateq} and $\eta(\vec r)=\eta$. As a consequence,
$\eta$ cancels in Eq.~\eqref{Tdef} and the obtained noise
temperatures are universal functions independent of the solvent
properties. All the subsequent results are derived under the latter approximation.

To gain a physical understanding of the functional form of
$\mathcal{T}(\omega)$, consider its origin from the hydrodynamic
coupling between the particle and distant solvent volume elements that
are locally equilibrated at different temperatures $T(\vec r)$. In our
low-Reynolds number approximation, the exchange of momentum is
dominated by vorticity diffusion \footnote{For rotation, Eq.~\eqref{sys_bis1}
  reduces to the diffusion equation $\partial_t \vec v= \nu \nabla^2
  \vec v$ for the velocity itself, since $\nabla p=0$ due to spherical
  symmetry. For translation, $\partial_t \vec v= -\nabla p/\varrho+
  \nu \nabla^2 \vec v$ can be understood as a diffusion equation where
  pressure acts as a source term.}  with the diffusivity given by the
kinematic viscosity $\nu\equiv  \eta/\varrho$.  This defines the
inverse characteristic time scale $\omega_{\text{f}} \equiv  2\nu/R^2$
for fluid transport over distances on the order of the particle
radius, as introduced in Eq.~\eqref{eq:rates}.

\emph{Low frequency fluctuations} are those with $\omega \ll
\omega_{\text{f}}$, during which the vorticity spreads out
considerably from the particle. Since the translational field is more
long-ranged than the rotational one ($\vec u^{\text{\tiny{T}}} \sim
1/r$ versus $\vec u^{\text{\tiny{R}}} \sim 1/r^2$), the translational
noise is effectively cooler, as it involves an average over
farther, i.e.\ cooler, regions of fluid.  Ultimately, in the limit
$\omega \to 0$, we find that the noise temperatures reduce to the
effective temperatures $T^{\scriptscriptstyle{X,\Theta}}_{\rm HBM}$ known to characterize
the overdamped hot Brownian motion of the positions and angles, respectively
\cite{Chakraborty.2011, Rings.2012}, for which we employ the
short-hand notation
\begin{align*}
& T^{\scriptscriptstyle X}\equiv  \mathcal{T}^{\text{\tiny{T}}}(0) =
T_0+\frac{5}{12}\Delta T \,,  \\ 
& T^{\scriptscriptstyle \Theta} \equiv 
\mathcal{T}^{\text{\tiny{R}}}(0)=T_0+\frac{3}{4}\Delta T\,.
\end{align*}

In contrast, during \emph{high-frequencies fluctuations} with $\omega \gg
\omega_{\text{f}}$, fluid momentum cannot diffuse significantly from
the particle surface. The vorticity emanating from a particle
oscillating at frequency $\omega$ cannot penetrate the fluid beyond
the skin depth $k_0^{-1}\equiv  (2\nu/\omega)^{1/2} \ll R$, resulting
in an exponential decay $\phi(r, \omega)\propto e^{-k_0(r-R)}$ of the
dissipation function; see Eqs.~\eqref{phi_transl}, \eqref{phi_rot}.
Therefore, the average in Eq.~\eqref{Tdef} is essentially restricted to a
thin skin of solvent around the particle surface, and the noise
temperatures tend towards the surface temperature $T_s=T_0+ \Delta T$ for large
$\omega$. However, note that the finite compressibility becomes relevant at very high frequencies. As a consequence, the noise temperature may deviate significantly from our predictions for frequencies larger than the inverse of the time it takes a sound wave in the solvent to traverse a distance $R$ \cite{Hinch.1993, Chow.1973}. \\
It is worth mentioning another peculiarity implied by the incompressibility assumption. To accelerate a particle in an incompressible fluid, the displaced fluid has to be moved from the
front to the back of the particle. Therefore, the layer of fluid that
is set into motion never collapses completely onto the particle
surface, as it does for rotation at high frequencies. While the noise temperature is not affected, since the bulk dissipation turns out to be subdominant (see Appendix~\ref{sec:hydro}), incompressibility results in a renormalized particle mass \cite{Landau.1987,Hauge.1973}:
\begin{equation}\label{eq:Meff}
M_{\text{eff}}=M+M\varrho/(2\varrho_\mathrm{p}).
\end{equation}
The added mass in Eq.~\eqref{eq:Meff}, owing to the inertia of the displaced fluid, becomes relevant in the following. In the next sections we analyze some
immediate implications of the above results for the dynamics of a hot
Brownian sphere that is either freely diffusing or trapped in a
confining potential.

\section{The kinetic temperature}\label{sec:Tkin}
The GLE's~\eqref{newton1},~\eqref{newton2} both contain a Gaussian
noise satisfying a fluctuation-dissipation relation with constant
effective temperatures in the high-frequency limit. Therefore, one may
expect to find Maxwell-Boltzmann distributions of translational and
angular velocities under stationary conditions, which is corroborated
by molecular dynamics simulations
\cite{Chakraborty.2011,Rings.2012}. We thus define the kinetic
temperatures such that the stationary averages of the velocities
satisfy
\begin{align}\label{def}
&\frac{3}{2} k_{\text{B}} T^{\scriptscriptstyle V}\equiv \frac{1}{2}M_{\text{eff}}\langle \vec V^2 \rangle,
&\frac{3}{2}k_{\text{B}} T^{\scriptscriptstyle \Omega}\equiv \frac{1}{2}I\langle \vec \Omega^2 \rangle, 
\end{align}
which reduce to the equipartition theorem with $T^{\scriptscriptstyle \Omega}=T^{\scriptscriptstyle V}=T_0$
in case of a constant fluid temperature $T(\vec r)\equiv  T_0$.  For
simplicity, we concentrate on the translational motion, in the
following, but the same procedure applies
also to the rotational motion. 

From the Fourier transform of Eq.~\eqref{GLEt} in the absence of
an external force,
\begin{equation}
-i \omega M\vec V(\omega) = -\zeta^{\text{\tiny{+}}}(\omega) \vec V(\omega) + \vec \xi^{\text{\tiny{T}}}(\omega)\,,
\end{equation}
we derive the velocity spectral density
\begin{equation}\label{spectral}
C_{V}(\omega)\equiv \langle \vec V(\omega) \cdot \vec V(-\omega)\rangle= |\mu(\omega)|^2 C_{\xi}^{\text{\tiny{T}}}(\omega) \,.
\end{equation}
Here 
\begin{equation}\label{C}
C_{\xi}^{\text{\tiny{T}}}(\omega)= 3 \kB \mathcal{T}^{\text{\tiny{T}}}(\omega) \zeta(\omega)
\end{equation}
is the noise spectral density and $\mu(\omega)$ is the particle mobility defined as
\begin{equation}\label{mu}
  \mu(\omega)=\frac{1}{\zeta^{\scriptscriptstyle+}(\omega)-i\omega M}\,.
\end{equation}
The Wiener-Khinchine theorem then gives the velocity auto-correlation function
\begin{equation}
\langle \vec V(t) \cdot \vec V(0) \rangle= \frac{1}{2 \pi}
\int_{-\infty}^{\infty} |\mu(\omega)|^2
C_{\xi}^\text{\tiny{T}}(\omega)e^{-i\omega t} \,\diff\omega \,,
\end{equation}
from which the translational kinetic temperature $T^{\scriptscriptstyle V}$, defined in
Eq.~\eqref{def}, follows as
\begin{equation}
T^{\scriptscriptstyle V}= \frac{M_{\text{eff}}}{\pi}\! \int_{0}^{\infty} \! |\mu(\omega)|^2 \mathcal{T}^{\text{\tiny{T}}}(\omega) \zeta(\omega) \,\diff\omega\,, \label{Tk}
\end{equation}
since the integrand is an even function of $\omega$.

To further evaluate this result, we introduce into the mobility $\mu$
the explicit expression for the memory kernel of a sphere translating
in an incompressible fluid with no-slip boundary conditions
\cite{Hauge.1973}:
\begin{equation}\label{zeta}
\zeta^\text{\tiny{+}}(\omega)=6 \pi \eta R \left[ 1+(1-i)\sqrt{\frac{R^2 \omega}{2\nu}}- iR^2\omega/9\nu\right]\,.
\end{equation}
The first term in the brackets is the usual Stokes friction $\zeta^\text{\tiny{+}}(\omega=0)\equiv \zeta$, the
second describes the vorticity diffusion and gives rise to the
long-time tails \cite{Zwanzig.1970, Fox.1983}. The third term accounts for the mentioned mass
renormalization, Eq.~\eqref{eq:Meff}. With the notation $x^2 =
\omega/\omega_{\text{f}}$, Eq.~\eqref{Tk} now reads
\begin{equation}
T^{\scriptscriptstyle V}=\frac{1}{\pi} \int_{0}^{\infty} \frac{4 \alpha x(x+1)\mathcal{T}^{\text{\tiny{T}}}(x)}{(1+x)^2+x^2(1+\alpha x)^2}\, \diff x\,, \label{TkFin}
\end{equation}
which depends on the particle-to-fluid density ratio via the parameter
$\alpha\equiv  2(2\varrho_\mathrm{p}/\varrho+1)/9$. The same procedure
gives the rotational kinetic temperature
\begin{widetext}
\begin{equation}\label{Tk2}
T^{\scriptscriptstyle \Omega} = \frac{1}{\pi} \int_{0}^{\infty}  \frac{12\beta x(1+2x+2x^2)(3+6x+6x^2+2x^3)  \mathcal{T}^{\text{\tiny{R}}}(x) }{[(3+6x+6x^2+2x^3)^2+ x^4(2(1+x)\!+\!3\beta(1+2x+2x^2))^2]} \diff x
\end{equation}
\end{widetext}
with $\beta \equiv  2\varrho_\mathrm{p}/(15\varrho)$.  

\begin{figure}
\begin{tikzpicture}
\begin{axis}[xmin=0,xmax=10,ymin=0.6 ,ymax=1.0,legend style={ at={(1,0.85)}, anchor=east}, xlabel=$\varrho_\mathrm{p}/\varrho$,
ylabel=$\left(T^{V,\Omega}-T_0\right)/\Delta T$] 
\addplot[color=red] coordinates {(0.001, 0.864392)(0.101, 0.851944)(0.201, 0.840953)(0.301, 
0.831111)(0.401, 0.822199)(0.501, 0.814058)(0.601, 0.806568) 
(0.701, 0.799633)(0.801, 0.79318)(0.901, 0.787148)(1.001, 
0.781488)(1.101, 0.776158)(1.201, 0.771123)(1.301, 0.766355) 
(1.401, 0.761827)(1.501, 0.757518)(1.601, 0.753409)(1.701, 
0.749483)(1.801, 0.745726)(1.901, 0.742125)(2.001, 0.738668) 
(2.101, 0.735346)(2.201, 0.732147)(2.301, 0.729065)(2.401, 
0.726092)(2.501, 0.723221)(2.601, 0.720446)(2.701, 0.71776) 
(2.801, 0.71516)(2.901, 0.71264)(3.001, 0.710195)(3.101, 
0.707823)(3.201, 0.705518)(3.301, 0.703278)(3.401, 0.701099) 
(3.501, 0.698979)(3.601, 0.696914)(3.701, 0.694902)(3.801, 
0.692941)(3.901, 0.691029)(4.001, 0.689162)(4.101, 0.68734) 
(4.201, 0.685561)(4.301, 0.683822)(4.401, 0.682123)(4.501, 
0.680461)(4.601, 0.678835)(4.701, 0.677243)(4.801, 0.675685) 
(4.901, 0.67416)(5.001, 0.672665)(5.101, 0.6712)(5.201, 
0.669765)(5.301, 0.668357)(5.401, 0.666976)(5.501, 0.665621) 
(5.601, 0.664291)(5.701, 0.662985)(5.801, 0.661703)(5.901, 
0.660444)(6.001, 0.659207)(6.101, 0.657992)(6.201, 0.656797) 
(6.301, 0.655622)(6.401, 0.654467)(6.501, 0.653331)(6.601, 
0.652214)(6.701, 0.651114)(6.801, 0.650032)(6.901, 0.648967) 
(7.001, 0.647918)(7.101, 0.646885)(7.201, 0.645868)(7.301, 
0.644867)(7.401, 0.64388)(7.501, 0.642907)(7.601, 0.641948) 
(7.701, 0.641004)(7.801, 0.640072)(7.901, 0.639154)(8.001, 
0.638248)(8.101, 0.637355)(8.201, 0.636474)(8.301, 0.635604) 
(8.401, 0.634747)(8.501, 0.6339)(8.601, 0.633065)(8.701, 
0.63224)(8.801, 0.631426)(8.901, 0.630622)(9.001, 0.629829) 
(9.101, 0.629045)(9.201, 0.628271)(9.301, 0.627506)(9.401, 
0.626751)(9.501, 0.626004)(9.601, 0.625267)(9.701, 0.624538) 
(9.801, 0.623818)(9.901, 0.623106)(10, 0.6224)};
\addplot[color=blue] coordinates {(0,1)(0.1, 0.971441)(0.2, 0.956735)(0.3, 0.945781)(0.4, 0.936945)(0.5, 0.929508)(0.6, 0.923076)(0.7, 0.917407)(0.8, 0.912339)  
(0.9, 0.907758)(1., 0.903582)(1.1, 0.899746)(1.2, 0.896201)  
(1.3, 0.892908)(1.4, 0.889836)(1.5, 0.886958)(1.6, 0.884253)  
(1.7, 0.881703)(1.8, 0.879292)(1.9, 0.877007)(2., 0.874836)  
(2.1, 0.872769)(2.2, 0.870799)(2.3, 0.868916)(2.4, 0.867114)  
(2.5, 0.865388)(2.6, 0.863732)(2.7, 0.86214)(2.8, 0.860609)  
(2.9, 0.859135)(3., 0.857713)(3.1, 0.856341)(3.2, 0.855016)  
(3.3, 0.853735)(3.4, 0.852496)(3.5, 0.851296)(3.6, 0.850132)  
(3.7, 0.849005)(3.8, 0.84791)(3.9, 0.846847)(4., 0.845814)  
(4.1, 0.84481)(4.2, 0.843834)(4.3, 0.842883)(4.4, 0.841957)  
(4.5, 0.841055)(4.6, 0.840176)(4.7, 0.839319)(4.8, 0.838483)  
(4.9, 0.837666)(5., 0.836869)(5.1, 0.83609)(5.2, 0.83533)  
(5.3, 0.834586)(5.4, 0.833858)(5.5, 0.833147)(5.6, 0.83245)  
(5.7, 0.831769)(5.8, 0.831101)(5.9, 0.830448)(6., 0.829807)  
(6.1, 0.82918)(6.2, 0.828564)(6.3, 0.827961)(6.4, 0.827369)  
(6.5, 0.826788)(6.6, 0.826218)(6.7, 0.825659)(6.8, 0.82511)  
(6.9, 0.824571)(7., 0.824041)(7.1, 0.823521)(7.2, 0.823009)  
(7.3, 0.822507)(7.4, 0.822013)(7.5, 0.821527)(7.6, 0.82105)  
(7.7, 0.82058)(7.8, 0.820117)(7.9, 0.819663)(8., 0.819215)  
(8.1, 0.818774)(8.2, 0.818341)(8.3, 0.817914)(8.4, 0.817493)  
(8.5, 0.817079)(8.6, 0.816671)(8.7, 0.816269)(8.8, 0.815873) 
(8.9, 0.815483)(9., 0.815098)(9.1, 0.814719)(9.2, 0.814345)
(9.3, 0.813976)(9.4, 0.813612)(9.5, 0.813254)(9.6, 0.8129) 
(9.7, 0.812551)(9.8, 0.812207)(9.9, 0.811867)(10., 0.811532)};
\legend{$T^{\scriptscriptstyle V}$,$T^{\scriptscriptstyle \Omega}$}
\end{axis}
\end{tikzpicture}
\caption{Rotational (\textcolor{blue}{---}) and translational (\textcolor{red}{---}) kinetic temperature as function of the density ratio $\varrho_\mathrm{p}/\varrho$.}
\label{fig:Tk}
\end{figure}
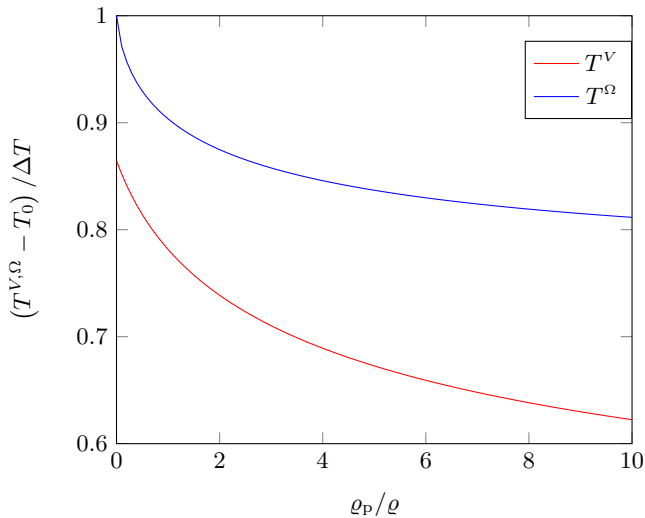

Equations \eqref{TkFin}, \eqref{Tk2} can be integrated numerically
using the translational and rotational noise temperatures
$\mathcal{T}(\omega)$ introduced in Section \ref{sec:Tomega}. The
results are shown in Figure \ref{fig:Tk}.  The kinetic temperatures
are seen to depend on the density ratio
$\varrho_\mathrm{p}/\varrho$. To understand this, consider a
translating sphere. In the Markov limit, its velocity relaxes within
the Stokes time, corresponding to the relaxation rate
\begin{equation}
\omega_\mathrm{p} \equiv  \frac{\zeta}{M}=
\frac{6\pi \eta R}{M} = \frac{9\varrho}{4\varrho_\mathrm{p}}\omega_\mathrm{f}\;,
\end{equation}  
introduced in Eq.~\eqref{eq:rates}.  The density ratio thus relates the
characteristic time for the kinematic equilibration of the particle with the fluid --- i.e. the time it takes to spread the particle momentum to a fluid mass comparable to the particle mass --- to the time it takes to spread its momentum to a fluid volume comparable to the particle volume. Accordingly, the kinematic equilibration affects either a small or large fluid volume compared to the
particle size, suggesting a kinetic temperature close to the
temperature $T_s$ at the particle surface or close to the stationary
effective temperature $T^{\scriptscriptstyle X}$, respectively.

Indeed, if $\varrho_\mathrm{p}/\varrho \ll 1$, only the upper part of
the spectrum $\mathcal{T}(\omega)$ contributes to the kinetic
temperatures, as seen from Eqs.~\eqref{TkFin},~\eqref{Tk2}, where the integrand
contributes significantly only for $x\! \gg \! 1$. Hence, the
rotational kinetic temperature $T^{\scriptscriptstyle \Omega}$ approaches the surface
temperature:
\begin{equation}
  T^{\scriptscriptstyle \Omega} \sim  \mathcal{T}^{\text{\tiny{R}}}(\infty)= T_0+\Delta
  T=T_s \; \text{ for } \varrho_\mathrm{p}/\varrho \to 0\;.
\end{equation}
Due to the mass renormalization, Eq.~\eqref{eq:Meff}, the
translational kinetic temperature $T^{\scriptscriptstyle V}$ always remains somewhat
below this limit, though. Although the noise temperature attempts to shake the
particle with a strength proportional to the surface temperature
$T_s$, the particle cannot move without exciting a long ranged flow
field that ultimately increases its own inertia. This effect limits the velocity fluctuations of
the particle to a non-universal apparent ``equipartition'' temperature
$T^{\scriptscriptstyle V}$ that depends on the density ratio $\varrho_\mathrm{p}/\varrho$,
and attains the limit
\begin{equation}
  T^{\scriptscriptstyle V}  \simeq T_0 + 0.86 \,\Delta T <T_\text{s} \; \text{ for } \varrho_\mathrm{p}/\varrho \to 0\;.
\end{equation}
As a consequence, the translational particle velocity never
thermalizes to the fluid temperature at the particle surface.

In the opposite limit, $\varrho_\mathrm{p}/\varrho \gg 1$, the
frequency-dependent terms in Eq.~\eqref{zeta}, which are proportional to
$R^2 \omega_\mathrm{p}/\nu = 2\omega_\mathrm{p}/ \omega_\mathrm{f}
\ll1$, become small. In this limit, the kinetic temperature approaches
the stationary values of the respective effective noise temperatures
$\mathcal{T}(0)$, which coincide with the known temperatures for the
configurational degrees of freedom, represented by the positional and
orientational coordinates $X$ and $\Theta$ \cite{Chakraborty.2011,
  Rings.2012} (see Sec.~\ref{sec:Tomega}). They determine the
translational and rotational diffusion coefficient of the hot Brownian
particle, e.g.\ for translation,
\begin{eqnarray}\label{DHBM}
  D&=& \lim_{t \to \infty}\frac{1}{6} \frac{\diff}{\diff t}\langle \left(\vec X(t) -\vec X(0)\right)^2 \rangle\\
  &=&\frac{1}{2}\int_{-\infty}^{\infty} \langle \vec V(t) \cdot \vec
  V(0) \rangle \diff t = \frac{1}{2} C_{V}(\omega)|_{\omega=0}\,. 
 \nonumber
\end{eqnarray}
Using Eqs.~\eqref{spectral}, \eqref{C}, \eqref{mu}, and \eqref{zeta},
we recover (to leading order in the temperature increment $\Delta T$,
i.e.\ not accounting for the temperature-induced spatial variations in
the viscosity) the generalized Einstein relation \cite{Chakraborty.2011}
\begin{equation}\label{Einstein}
D_{\text{HBM}}=\frac{\kB \mathcal{T}^{\text{\tiny{T}}}(\omega)}{\zeta^\text{\tiny{+}}(\omega)}\Bigg |_{\omega=0}=\frac{\kB (T_0+\frac{5}{12}\Delta T)}{6 \pi \eta R}\;.
\end{equation}
The same reasoning applies to the orientation $\Theta$. Hence, we see
that for a hot Brownian particle that is much denser than the solvent,
the kinetic temperatures reduce to the effective configurational
temperatures,
\begin{equation}
T^{\scriptscriptstyle{V,\Omega}} \sim  \mathcal{T^{\text{\tiny{T}},\text{\tiny{R}}}}(0) =
T^{\scriptscriptstyle{X,\Theta}}  \; \text{ for } \varrho_\mathrm{p}/\varrho \to \infty\,.
\end{equation}

Moreover, in any case, both the translational and rotational
velocities of a hot spherical particle can be statistically
characterized by a (non-universal) Maxwell--Boltzmann distribution
\begin{equation}\label{Boltzmann}
P(\vec V,\vec \Omega) \propto \exp \left(- \frac{M_\text{eff} \vec
    V^2}{2 \kB T^{\scriptscriptstyle V}} - \frac{I \vec
    \Omega^2}{2 \kB T^{\scriptscriptstyle \Omega}}\right) 
\end{equation}
with effective temperatures that depend on the density ratio
$\varrho_\mathrm{p}/\varrho$, in agreement with the fact that probability distributions of non-equilibrium ensembles explicitly depend on the dynamics of the system.

\section{Particle in a harmonic potential}\label{sec:pos}

The discussion of the previous section can be repeated for a particle
trapped in a harmonic potential. While the kinetic temperature of a
free particle is determined by the competition between the vorticity
diffusion time $\omega_{\text{f}}^{-1}$ and the Stokes relaxation time
$\omega_{\text{p}}^{-1}$ introduced in Eq.~\eqref{eq:rates}, a
sufficiently narrow confining potential introduces an additional
interfering time scale. In the following, we examine more closely the
case of translational diffusion in confinement, but qualitatively
similar results can be derived for the rotational case. 

\begin{figure}
\begin{tikzpicture}
  \begin{axis}[xmin=0,xmax=10,ymin=0.6 ,ymax=1.0,legend style={
      at={(axis cs: 10,1)}, anchor=north east},
    xlabel=$\varrho_\mathrm{p}/\varrho$,
    ylabel=$\left(T^{\scriptscriptstyle V}-T_0\right)/\Delta T$]
    \addplot[color=red] coordinates
    {(0.01,0.857664)(0.11,0.846268)(0.21,0.836042)(0.31,0.826781)(0.41,0.818327)(0.51,0.810557)(0.61,0.803372)(0.71,0.796694)(0.81,0.79046)(0.91,0.784616)(1.01,0.77912)(1.11,0.773934)(1.21,0.769027)(1.31,0.764372)(1.41,0.759947)(1.51,0.75573)(1.61,0.751705)(1.71,0.747856)(1.81,0.744169)(1.91,0.740633)(2.01,0.737235)(2.11,0.733967)(2.21,0.730819)(2.31,0.727785)(2.41,0.724856)(2.51,0.722026)(2.61,0.719289)(2.71,0.71664)(2.81,0.714074)(2.91,0.711586)(3.01,0.709171)(3.11,0.706827)(3.21,0.70455)(3.31,0.702335)(3.41,0.700181)(3.51,0.698084)(3.61,0.696041)(3.71,0.69405)(3.81,0.692109)(3.91,0.690216)(4.01,0.688368)(4.11,0.686564)(4.21,0.684801)(4.31,0.683078)(4.41,0.681394)(4.51,0.679747)(4.61,0.678135)(4.71,0.676558)(4.81,0.675013)(4.91,0.6735)(5.01,0.672018)(5.11,0.670565)(5.21,0.66914)(5.31,0.667743)(5.41,0.666373)(5.51,0.665028)(5.61,0.663708)(5.71,0.662412)(5.81,0.66114)(5.91,0.65989)(6.01,0.658661)(6.11,0.657454)(6.21,0.656268)(6.31,0.655101)(6.41,0.653954)(6.51,0.652826)(6.61,0.651715)(6.71,0.650623)(6.81,0.649548)(6.91,0.648489)(7.01,0.647447)(7.11,0.646421)(7.21,0.64541)(7.31,0.644414)(7.41,0.643433)(7.51,0.642466)(7.61,0.641513)(7.71,0.640574)(7.81,0.639648)(7.91,0.638735)(8.01,0.637834)(8.11,0.636946)(8.21,0.636069)(8.31,0.635205)(8.41,0.634352)(8.51,0.63351)(8.61,0.632679)(8.71,0.631859)(8.81,0.631049)(8.91,0.630249)(9.01,0.62946)(9.11,0.62868)(9.21,0.62791)(9.31,0.627149)(9.41,0.626397)(9.51,0.625654)(9.61,0.62492)(9.71,0.624195)(9.81,0.623478)(9.91,0.62277)(10.01,0.622069)
    }; \addplot[color=blue] coordinates
    {(0.01,0.858914)(0.11,0.847732)(0.21,0.837712)(0.31,0.82865)(0.41,0.820389)(0.51,0.812805)(0.61,0.805801)(0.71,0.799299)(0.81,0.793235)(0.91,0.787558)(1.01,0.782224)(1.11,0.777196)(1.21,0.772443)(1.31,0.767939)(1.41,0.763661)(1.51,0.759588)(1.61,0.755704)(1.71,0.751992)(1.81,0.74844)(1.91,0.745035)(2.01,0.741767)(2.11,0.738626)(2.21,0.735603)(2.31,0.732691)(2.41,0.729882)(2.51,0.727169)(2.61,0.724548)(2.71,0.722013)(2.81,0.719558)(2.91,0.71718)(3.01,0.714874)(3.11,0.712636)(3.21,0.710463)(3.31,0.708352)(3.41,0.706299)(3.51,0.704302)(3.61,0.702357)(3.71,0.700464)(3.81,0.698618)(3.91,0.696819)(4.01,0.695064)(4.11,0.693351)(4.21,0.691679)(4.31,0.690045)(4.41,0.688449)(4.51,0.686888)(4.61,0.685362)(4.71,0.683869)(4.81,0.682407)(4.91,0.680977)(5.01,0.679576)(5.11,0.678203)(5.21,0.676858)(5.31,0.675539)(5.41,0.674246)(5.51,0.672978)(5.61,0.671734)(5.71,0.670512)(5.81,0.669314)(5.91,0.668137)(6.01,0.66698)(6.11,0.665845)(6.21,0.664729)(6.31,0.663632)(6.41,0.662554)(6.51,0.661494)(6.61,0.660451)(6.71,0.659425)(6.81,0.658416)(6.91,0.657423)(7.01,0.656446)(7.11,0.655484)(7.21,0.654537)(7.31,0.653604)(7.41,0.652685)(7.51,0.65178)(7.61,0.650888)(7.71,0.650009)(7.81,0.649143)(7.91,0.648289)(8.01,0.647447)(8.11,0.646617)(8.21,0.645798)(8.31,0.644991)(8.41,0.644194)(8.51,0.643408)(8.61,0.642633)(8.71,0.641868)(8.81,0.641113)(8.91,0.640367)(9.01,0.639631)(9.11,0.638904)(9.21,0.638187)(9.31,0.637478)(9.41,0.636778)(9.51,0.636087)(9.61,0.635404)(9.71,0.634729)(9.81,0.634062)(9.91,0.633403)(10.01,0.632752)
    }; \addplot[color=mygreen] coordinates
    {(0.01,0.874741)(0.11,0.86625)(0.21,0.858673)(0.31,0.851853)(0.41,0.845728)(0.51,0.840012)(0.61,0.834815)(0.71,0.830009)(0.81,0.825545)(0.91,0.821379)(1.01,0.817477)(1.11,0.81381)(1.21,0.810353)(1.31,0.807084)(1.41,0.803986)(1.51,0.801042)(1.61,0.798239)(1.71,0.795565)(1.81,0.79301)(1.91,0.790563)(2.01,0.788218)(2.11,0.785965)(2.21,0.7838)(2.31,0.781715)(2.41,0.779705)(2.51,0.777766)(2.61,0.775893)(2.71,0.774082)(2.81,0.772329)(2.91,0.770631)(3.01,0.768985)(3.11,0.767388)(3.21,0.765837)(3.31,0.764331)(3.41,0.762865)(3.51,0.761439)(3.61,0.760051)(3.71,0.758699)(3.81,0.757382)(3.91,0.756097)(4.01,0.754844)(4.11,0.753622)(4.21,0.752429)(4.31,0.751264)(4.41,0.750125)(4.51,0.749013)(4.61,0.747926)(4.71,0.746862)(4.81,0.745784)(4.91,0.7448)(5.01,0.7437)(5.11,0.7426)(5.21,0.7415)(5.31,0.740852)(5.41,0.739922)(5.51,0.739009)(5.61,0.738113)(5.71,0.737233)(5.81,0.736368)(5.91,0.735519)(6.01,0.734684)(6.11,0.733863)(6.21,0.733056)(6.31,0.732262)(6.41,0.731481)(6.51,0.730713)(6.61,0.729957)(6.71,0.729212)(6.81,0.728479)(6.91,0.727757)(7.01,0.727046)(7.11,0.726345)(7.21,0.725655)(7.31,0.724975)(7.41,0.724304)(7.51,0.723643)(7.61,0.722991)(7.71,0.722348)(7.81,0.721714)(7.91,0.721088)(8.01,0.72047)(8.11,0.719861)(8.21,0.71926)(8.31,0.718666)(8.41,0.71808)(8.51,0.717501)(8.61,0.71693)(8.71,0.716365)(8.81,0.715808)(8.91,0.715257)(9.01,0.714713)(9.11,0.714175)(9.21,0.713643)(9.31,0.713118)(9.41,0.712598)(9.51,0.712085)(9.61,0.711577)(9.71,0.711075)(9.81,0.710579)(9.91,0.710088)(10.01,0.709602)
    }; \addplot[color=myyellow] coordinates
    {(0.01,0.898171)(0.11,0.892021)(0.21,0.886565)(0.31,0.881675)(0.41,0.877252)(0.51,0.873218)(0.61,0.869514)(0.71,0.866091)(0.81,0.862912)(0.91,0.859944)(1.01,0.857164)(1.11,0.854548)(1.21,0.852079)(1.31,0.849746)(1.41,0.847525)(1.51,0.845428)(1.61,0.843406)(1.71,0.841482)(1.81,0.839641)(1.91,0.837877)(2.01,0.836182)(2.11,0.834552)(2.21,0.832982)(2.31,0.831467)(2.41,0.830005)(2.51,0.828591)(2.61,0.827223)(2.71,0.825898)(2.81,0.824613)(2.91,0.823366)(3.01,0.822154)(3.11,0.820977)(3.21,0.819832)(3.31,0.818717)(3.41,0.817631)(3.51,0.816573)(3.61,0.81554)(3.71,0.814533)(3.81,0.813549)(3.91,0.812588)(4.01,0.811649)(4.11,0.81073)(4.21,0.809832)(4.31,0.808952)(4.41,0.80809)(4.51,0.807246)(4.61,0.806419)(4.71,0.805608)(4.81,0.804813)(4.91,0.804033)(5.01,0.803267)(5.11,0.802515)(5.21,0.801777)(5.31,0.801051)(5.41,0.800339)(5.51,0.799636)(5.61,0.798949)(5.71,0.798271)(5.81,0.797605)(5.91,0.796949)(6.01,0.796304)(6.11,0.795668)(6.21,0.795043)(6.31,0.794426)(6.41,0.793818)(6.51,0.793219)(6.61,0.792629)(6.71,0.792048)(6.81,0.791477)(6.91,0.790917)(7.01,0.79036)(7.11,0.789808)(7.21,0.789257)(7.31,0.788719)(7.41,0.78819)(7.51,0.787675)(7.61,0.787166)(7.71,0.786662)(7.81,0.78616)(7.91,0.785659)(8.01,0.78516)(8.11,0.784667)(8.21,0.784184)(8.31,0.783717)(8.41,0.78326)(8.51,0.782807)(8.61,0.782356)(8.71,0.781902)(8.81,0.781446)(8.91,0.780993)(9.01,0.780548)(9.11,0.780112)(9.21,0.779689)(9.31,0.779274)(9.41,0.778863)(9.51,0.778453)(9.61,0.778042)(9.71,0.777632)(9.81,0.777224)(9.91,0.776821)(10.01,0.776425)
    }; \legend{$10^{-3}$,$10^{-1}$,
      $10$,
      $10^2$}
\end{axis}
\end{tikzpicture}
\caption{Kinetic temperature $T^{\scriptscriptstyle V}$ of a
  particle in harmonic confinement, as given by Eq.~\eqref{Tkh} for various $\omega_\text{t}/\omega_\text{f}= 10^{-3},...,10^2$. For
  $\omega_\text{t}/\omega_\text{f}, \lesssim 10^{-1}$,
  $T^{\scriptscriptstyle V}$ is hardly distinguishable from the
  kinetic temperature of a free particle.}
\label{fig:Tkh}
\end{figure}
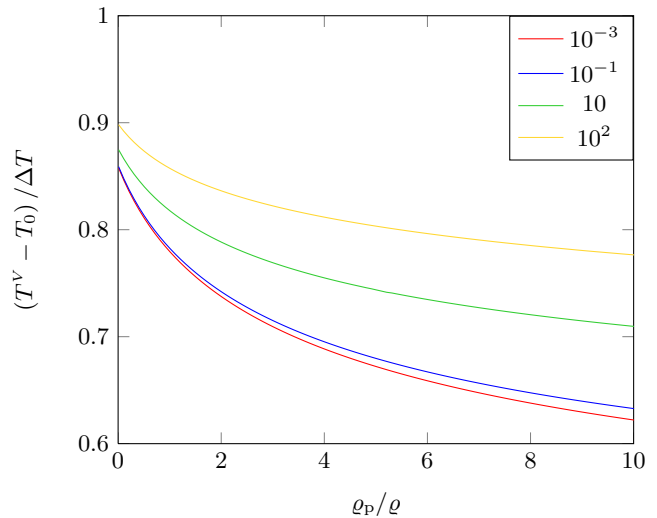

The parabolic confinement potential $\mathcal{U}(\vec X)=K\vec X^2/2$
gives rise to the trap relaxation time
\begin{equation}
\omega_{\text{t}}^{-1} =
6 \pi \eta R/K= \omega_\text{p}/\omega_0^2  \;,
\end{equation} 
where $\omega_{\text{0}}^2=K/M$ is the undamped oscillation frequency.
With $\vec F_{\text{e}}=-K\vec X$, the Fourier transformed Eq.~\eqref{GLEt},
\begin{equation*} 
-M\omega^2\vec X(\omega)=i\omega\zeta^{\scriptscriptstyle+}(\omega)\vec X(\omega) - K\vec X(\omega) + \xi(\omega)\,,
\end{equation*}
yields the spectral density
\begin{equation}\label{spectral_pos}
C_{X}(\omega)\equiv \langle \vec X(\omega) \cdot \vec X(-\omega)\rangle= |R(\omega)|^2 C_{\xi}(\omega) \,,
\end{equation}
where the (positional) response function is defined by
\begin{equation}\label{R}
R(\omega)=\frac{1}{M(\omega_0^2-\omega^2) -i\omega\zeta^\text{\tiny{+}}(\omega)}.
\end{equation}
We use the relation 
\begin{equation*}
\langle \vec V(t) \cdot \vec V(0) \rangle=-\frac{\diff^2 }{\diff t^2}\langle \vec X(t) \cdot \vec X(0) \rangle\,,
\end{equation*}
between the stationary correlation functions for position and velocity
in frequency space, $C_{V}(\omega)= \omega^2 C_{X}(\omega)$. The
kinetic temperature, as defined in Eq.~\eqref{def}, follows as
\begin{eqnarray}\nonumber
T^{\scriptscriptstyle V}&=& \frac{M_{\text{eff}}}{\pi}\! \int_{0}^{\infty} \! \omega^2|R(\omega)|^2 \mathcal{T}^{\text{\tiny{T}}}(\omega) \zeta(\omega) \,\diff\omega \\
&=&\frac{1}{\pi} \int_{0}^{\infty} \!\!\!\frac{4 \alpha
  x^5(x+1)\mathcal{T}^{\text{\tiny{T}}}(x)}{x^4(1+x)^2+(x^3+\alpha
  x^4 - \omega_\text{t}/\omega_\text{f} )^2}\, \diff x. \;\;\;  \label{Tkh}
\end{eqnarray}
The result is again integrated numerically and depicted in
Fig.~\ref{fig:Tkh}.  Clearly, if $\omega_\text{t} \ll
\omega_\text{f}$, which means that the potential is not effective
while the velocity is relaxing, we recover the result for free
diffusion, Eq.~\eqref{TkFin}. This should be the case for an optically trapped nano-particle in water under standard experimental conditions. Indeed
for a gold particle with $R\simeq100\,\text{nm}$, assuming a trap stiffness
$K=10^{-6} \text{Nm}^{-1}$ \cite{Hansen.2005}, we estimate
$\omega_\text{t}/\omega_\text{f}\simeq10^{-2}$. The velocity
relaxation time decreases as we increase the ratio
$\omega_\text{t}/\omega_\text{f}$, resulting in a higher kinetic
temperature. When $\omega_\text{t} \gg \omega_\text{f}$ the narrow
confinement eventually overrides the inertia of the particle motion
due to its effective mass $M_\text{eff}$, so that the kinetic
temperature $T^{\scriptscriptstyle V}$ approaches the surface temperature $T_s$.

Similar as for the particle velocity, we define the positional
temperature of a hot Brownian particle in a harmonic potential via the
generalized equipartition theorem,
\begin{equation}
\frac{3}{2} \kB T^{\scriptscriptstyle X} \equiv \frac{1}{2}\omega_0^2 M \langle \vec X^2 \rangle\,,
\end{equation}
where the average is taken with respect to the stationary
distribution. Using Eqs.~\eqref{spectral_pos} and \eqref{R} we
straightforwardly obtain
\begin{eqnarray}\nonumber
T^{\scriptscriptstyle X}&=& \frac{\omega_0^2 M }{\pi}\! \int_{0}^{\infty} \! |R(\omega)|^2 \mathcal{T}^{\text{\tiny{T}}}(\omega) \zeta(\omega) \,\diff\omega \\
&=&\frac{1}{\pi} \int_{0}^{\infty} \!\!\!\frac{4 (\omega_\text{t}/\omega_\text{f}) x (x+1)\mathcal{T}^{\text{\tiny{T}}}(x)}{x^4(1+x)^2+(x^3+\alpha x^4 - \omega_\text{t}/\omega_\text{f} )^2}\, \diff x. \;\;\;\; \label{thbm_trap}
\end{eqnarray}
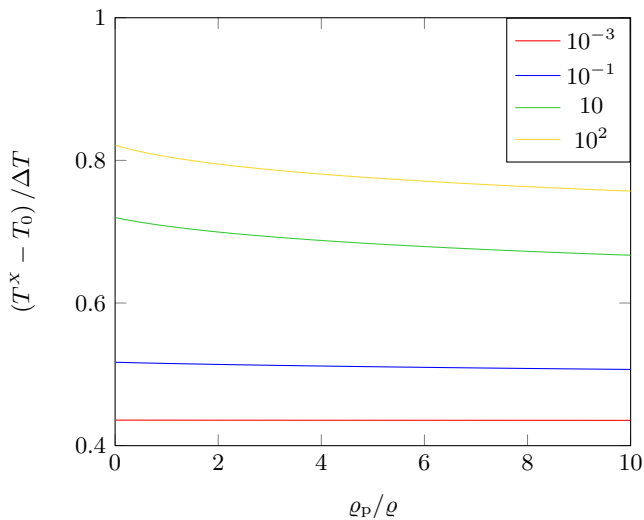
\begin{figure}
\begin{tikzpicture}
  \begin{axis}[xmin=0,xmax=10,ymin=0.4,ymax=1.0,legend style={
      at={(axis cs: 10,1)}, anchor=north east},
    xlabel=$\varrho_\mathrm{p}/\varrho$,
    ylabel=$\left(T^{\scriptscriptstyle X}-T_0\right)/\Delta
    T$]
    \addplot[color=red] coordinates
    {(0.01,0.435807)(0.11,0.435803)(0.21,0.435799)(0.31,0.435794)(0.41,0.43579)(0.51,0.435786)(0.61,0.435782)(0.71,0.435778)(0.81,0.435774)(0.91,0.43577)(1.01,0.435766)(1.11,0.435763)(1.21,0.435759)(1.31,0.435755)(1.41,0.435751)(1.51,0.435748)(1.61,0.435744)(1.71,0.43574)(1.81,0.435737)(1.91,0.435733)(2.01,0.43573)(2.11,0.435726)(2.21,0.435723)(2.31,0.435719)(2.41,0.435716)(2.51,0.435712)(2.61,0.435709)(2.71,0.435705)(2.81,0.435702)(2.91,0.435699)(3.01,0.435695)(3.11,0.435692)(3.21,0.435689)(3.31,0.435686)(3.41,0.435682)(3.51,0.435679)(3.61,0.435676)(3.71,0.435673)(3.81,0.43567)(3.91,0.435667)(4.01,0.435663)(4.11,0.43566)(4.21,0.435657)(4.31,0.435654)(4.41,0.435651)(4.51,0.435648)(4.61,0.435645)(4.71,0.435642)(4.81,0.435639)(4.91,0.435636)(5.01,0.435633)(5.11,0.43563)(5.21,0.435627)(5.31,0.435624)(5.41,0.435621)(5.51,0.435619)(5.61,0.435616)(5.71,0.435613)(5.81,0.43561)(5.91,0.435607)(6.01,0.435604)(6.11,0.435601)(6.21,0.435599)(6.31,0.435596)(6.41,0.435593)(6.51,0.43559)(6.61,0.435588)(6.71,0.435585)(6.81,0.435582)(6.91,0.435579)(7.01,0.435577)(7.11,0.435574)(7.21,0.435571)(7.31,0.435569)(7.41,0.435566)(7.51,0.435563)(7.61,0.435561)(7.71,0.435558)(7.81,0.435555)(7.91,0.435553)(8.01,0.43555)(8.11,0.435548)(8.21,0.435545)(8.31,0.435542)(8.41,0.43554)(8.51,0.435537)(8.61,0.435535)(8.71,0.435532)(8.81,0.43553)(8.91,0.435527)(9.01,0.435525)(9.11,0.435522)(9.21,0.43552)(9.31,0.435517)(9.41,0.435515)(9.51,0.435512)(9.61,0.43551)(9.71,0.435507)(9.81,0.435505)(9.91,0.435502)(10,0.43550)};
    \addplot[color=blue] coordinates
    {(0.01,0.516922)(0.11,0.51674)(0.21,0.516563)(0.31,0.516391)(0.41,0.516222)(0.51,0.516056)(0.61,0.515894)(0.71,0.515736)(0.81,0.51558)(0.91,0.515427)(1.01,0.515277)(1.11,0.51513)(1.21,0.514985)(1.31,0.514842)(1.41,0.514702)(1.51,0.514564)(1.61,0.514428)(1.71,0.514294)(1.81,0.514163)(1.91,0.514033)(2.01,0.513904)(2.11,0.513778)(2.21,0.513653)(2.31,0.51353)(2.41,0.513409)(2.51,0.513289)(2.61,0.51317)(2.71,0.513053)(2.81,0.512938)(2.91,0.512823)(3.01,0.512711)(3.11,0.512599)(3.21,0.512489)(3.31,0.51238)(3.41,0.512272)(3.51,0.512165)(3.61,0.512059)(3.71,0.511955)(3.81,0.511852)(3.91,0.511749)(4.01,0.511648)(4.11,0.511548)(4.21,0.511448)(4.31,0.51135)(4.41,0.511253)(4.51,0.511156)(4.61,0.511061)(4.71,0.510966)(4.81,0.510872)(4.91,0.510779)(5.01,0.510687)(5.11,0.510596)(5.21,0.510505)(5.31,0.510416)(5.41,0.510327)(5.51,0.510238)(5.61,0.510151)(5.71,0.510064)(5.81,0.509978)(5.91,0.509893)(6.01,0.509808)(6.11,0.509724)(6.21,0.509641)(6.31,0.509558)(6.41,0.509476)(6.51,0.509395)(6.61,0.509314)(6.71,0.509234)(6.81,0.509154)(6.91,0.509075)(7.01,0.508997)(7.11,0.508919)(7.21,0.508842)(7.31,0.508765)(7.41,0.508689)(7.51,0.508613)(7.61,0.508538)(7.71,0.508464)(7.81,0.50839)(7.91,0.508316)(8.01,0.508243)(8.11,0.50817)(8.21,0.508098)(8.31,0.508027)(8.41,0.507956)(8.51,0.507885)(8.61,0.507815)(8.71,0.507745)(8.81,0.507676)(8.91,0.507607)(9.01,0.507538)(9.11,0.50747)(9.21,0.507402)(9.31,0.507335)(9.41,0.507268)(9.51,0.507202)(9.61,0.507136)(9.71,0.50707)(9.81,0.507005)(9.91,0.50694)(10,0.50690)};
    \addplot[color=mygreen] coordinates
    {(0.01,0.719708)(0.11,0.718245)(0.21,0.716862)(0.31,0.715549)(0.41,0.714298)(0.51,0.713103)(0.61,0.711958)(0.71,0.710859)(0.81,0.709803)(0.91,0.708785)(1.01,0.707802)(1.11,0.706853)(1.21,0.705934)(1.31,0.705044)(1.41,0.70418)(1.51,0.703341)(1.61,0.702526)(1.71,0.701733)(1.81,0.700961)(1.91,0.700209)(2.01,0.699475)(2.11,0.698759)(2.21,0.69806)(2.31,0.697377)(2.41,0.696709)(2.51,0.696056)(2.61,0.695417)(2.71,0.694791)(2.81,0.694177)(2.91,0.693576)(3.01,0.692987)(3.11,0.692408)(3.21,0.691841)(3.31,0.691284)(3.41,0.690737)(3.51,0.6902)(3.61,0.689671)(3.71,0.689152)(3.81,0.688642)(3.91,0.68814)(4.01,0.687645)(4.11,0.687159)(4.21,0.68668)(4.31,0.686209)(4.41,0.685745)(4.51,0.685287)(4.61,0.684837)(4.71,0.684392)(4.81,0.683954)(4.91,0.683522)(5.01,0.683096)(5.11,0.682676)(5.21,0.682262)(5.31,0.681853)(5.41,0.681449)(5.51,0.68105)(5.61,0.680657)(5.71,0.680268)(5.81,0.679884)(5.91,0.679505)(6.01,0.67913)(6.11,0.67876)(6.21,0.678394)(6.31,0.678033)(6.41,0.677675)(6.51,0.677322)(6.61,0.676972)(6.71,0.676627)(6.81,0.676285)(6.91,0.675947)(7.01,0.675612)(7.11,0.675281)(7.21,0.674954)(7.31,0.67463)(7.41,0.674309)(7.51,0.673991)(7.61,0.673677)(7.71,0.673366)(7.81,0.673057)(7.91,0.672752)(8.01,0.67245)(8.11,0.672151)(8.21,0.671854)(8.31,0.671561)(8.41,0.67127)(8.51,0.670981)(8.61,0.670695)(8.71,0.670412)(8.81,0.670132)(8.91,0.669854)(9.01,0.669578)(9.11,0.669305)(9.21,0.669034)(9.31,0.668765)(9.41,0.668499)(9.51,0.668235)(9.61,0.667973)(9.71,0.667715)(9.81,0.66746)(9.91,0.667208)(10,0.66708)};
    \addplot[color=myyellow] coordinates
    {(0.01,0.821137)(0.11,0.819058)(0.21,0.817128)(0.31,0.815326)(0.41,0.813633)(0.51,0.812035)(0.61,0.810521)(0.71,0.809082)(0.81,0.80771)(0.91,0.806399)(1.01,0.805143)(1.11,0.803927)(1.21,0.802777)(1.31,0.801579)(1.41,0.80058)(1.51,0.799538)(1.61,0.798529)(1.71,0.797552)(1.81,0.796604)(1.91,0.795683)(2.01,0.794788)(2.11,0.793919)(2.21,0.793072)(2.31,0.792247)(2.41,0.79144)(2.51,0.790655)(2.61,0.789889)(2.71,0.789142)(2.81,0.78841)(2.91,0.787693)(3.01,0.786989)(3.11,0.786305)(3.21,0.785633)(3.31,0.784974)(3.41,0.784328)(3.51,0.783695)(3.61,0.783073)(3.71,0.782462)(3.81,0.781863)(3.91,0.781274)(4.01,0.780695)(4.11,0.780126)(4.21,0.779567)(4.31,0.779017)(4.41,0.778475)(4.51,0.777942)(4.61,0.777417)(4.71,0.776901)(4.81,0.776392)(4.91,0.775891)(5.01,0.775397)(5.11,0.77491)(5.21,0.77443)(5.31,0.773955)(5.41,0.773489)(5.51,0.773028)(5.61,0.772574)(5.71,0.772125)(5.81,0.771682)(5.91,0.771246)(6.01,0.770814)(6.11,0.770388)(6.21,0.769966)(6.31,0.76955)(6.41,0.769139)(6.51,0.768733)(6.61,0.768333)(6.71,0.767936)(6.81,0.767543)(6.91,0.767155)(7.01,0.766771)(7.11,0.766391)(7.21,0.766017)(7.31,0.765646)(7.41,0.765278)(7.51,0.764915)(7.61,0.764555)(7.71,0.764198)(7.81,0.763846)(7.91,0.763497)(8.01,0.763152)(8.11,0.76281)(8.21,0.762471)(8.31,0.762136)(8.41,0.761803)(8.51,0.761474)(8.61,0.761148)(8.71,0.760824)(8.81,0.760504)(8.91,0.760187)(9.01,0.759874)(9.11,0.759563)(9.21,0.759256)(9.31,0.75895)(9.41,0.758645)(9.51,0.75834)(9.61,0.758038)(9.71,0.757739)(9.81,0.75745)(9.91,0.757165)(10,0.7569)
    }; \legend{$
   10^{-3}$,$10^{-1}$,
      $10$,
      $10^2$}
\end{axis}
\end{tikzpicture}
\caption{Positional temperature $T^{\scriptscriptstyle X}$ of
  a particle in harmonic confinement, as given by
  Eq.~\eqref{thbm_trap} for various $\omega_\text{t}/\omega_\text{f}= 10^{-3},...,10^2$. At small $\omega_\text{t}/\omega_\text{f}$ the
  temperature is independent of $\varrho_\mathrm{p}/\varrho$ .}
\label{fig:Th}
\end{figure}

This result is integrated numerically and plotted in
Fig.~\ref{fig:Th}.  Again, if $\omega_\text{t} \ll \omega_\text{f}$,
we recover the configurational temperature of a free particle, since
the integrand in Eq.~\eqref{thbm_trap} is sharply peaked at $x\ll1$,
corresponding to $\omega\ll \omega_\text{f}$. Physically, the
relaxation in the potential takes place quasi-statically with respect
to the free hot Brownian motion, which can then be represented in the
Markov approximation, in perfect analogy to the equilibrium case. The
corresponding Langevin equation is
\begin{align*}
&\zeta\dot{\vec X}\!=\! -\nabla \mathcal{U} + \vec \xi,  &&\langle \xi_i(t) \xi_j (t')\rangle\! =\! 2D_{\text{HBM}}\delta(t-t')\delta_{ij},
\end{align*}
and its stationary solution is the generalized Boltzmann distribution
\begin{equation*}
P(\vec X) \propto \exp\left(-\frac{\mathcal{U}(\vec X)}{\kB T^{\scriptscriptstyle X}}\right)\,.
\end{equation*}
with the effective temperature 
\begin{equation}
  T^{\scriptscriptstyle X} = \zeta D_{\text{HBM}}= T_0+\frac{5}{12}\Delta T
\end{equation} 
\ \\
of free hot Brownian motion \cite{Chakraborty.2011}  (originally
denoted by $\thbm$).

In contrast, if $\omega_\text{t} \approx \omega_\text{p}$ the
potential interferes with the relaxation of the particle, resulting in
a higher $T^{\scriptscriptstyle X}$ than in the free case. Eventually, in the extreme limit
$\omega_\text{t} \gg \omega_\text{p}$, the integral peaks near
$\omega_\text{t}$, and $T^{\scriptscriptstyle X}$ approaches the kinetic temperature $T^{\scriptscriptstyle V}$
(non-uniformly in $\varrho_\mathrm{p}/\varrho$).  It is moreover worth
noting that the stationary probability distribution can in any case
still be written in the form of Eq.~\eqref{Boltzmann}, albeit with
non-universal temperatures $T^{\scriptscriptstyle V}$ and $T^{\scriptscriptstyle X}$ that generally depend on
the density ratio $\varrho_\mathrm{p}/\varrho$ and on the stiffness
$K$ of the potential. Analogous conclusions hold for the rotational
degrees of freedom.

\section{Conclusion}\label{sec:conclusion}
Starting from the fluctuating hydrodynamic description of the solvent,
which we required to be in local thermal equilibrium with an
inhomogeneous temperature field $T(\vec r)$, we have derived a
generalized Langevin equation for the motion of a suspended
particle. While the discussion was limited to the important case of
hot Brownian motion, where $T(\vec r)$ decays radially around the
particle, essentially the same reasoning applies to more general
temperature profiles \cite{Falasco.2014}. As a consequence of the non-isothermal
conditions, the noise temperature $\mathcal{T}(\omega)$ characterizing
the strength of the stochastic Langevin forces becomes frequency
dependent and differs for different degrees of freedom, which couple
to different hydrodynamic modes. From the noise temperature, we
derived approximate expressions for the effective temperatures at
which the rotational and translational degrees of freedom of a
spherical particle appear to thermalize. Explicit numerical results
have been limited to first order in the temperature increment $\Delta
T$, so that the temperature-dependence of the fluid viscosity could be
neglected. We found the (angular) velocities to be Maxwell--Boltzmann
distributed with non-universal, but explicitly known, effective temperatures. In the
long-time limit we regained previous results for the configurational
temperatures governing free and weakly confined hot Brownian motion.
 
\begin{widetext}
\appendix
\section{Derivation of the GLE's noise autocorrelation function}\label{sec:derivation}
Extending the calculation presented in \cite{Bedaux.1974} to a
non-isothermal solvent, we derive the expressions Eq.~\eqref{result1} and
Eq.~\eqref{result2} of Section \ref{sec:fh} for the translational motion. The same procedure can be applied separately to rotational motion
bearing in mind that, tanks to linearity and spherical symmetry, the
flow field $\vec v$ can be diveded into the two independent fields
$\vec {v}^{\text{\tiny{T}}}$ and $\vec {v}^{\text{\tiny{R}}}$
generated, respectively, by the particle translation and rotation, and
satisfying the boundary conditions:
\begin{eqnarray*}
&&\vec v^{\text{\tiny{T}}}(\vec r, t)= \vec V(t) \, \text{ on } \mathcal{S}\, ,\\
&&\vec v^{\text{\tiny{R}}}(\vec r, t)= \vec \Omega(t) \times \vec r\, \text{ on } \mathcal{S} .
\end{eqnarray*}
Since we focus on the translational motion only, we omit the superscript T. Using Eq.~\eqref{sysforce1}, the Fourier transform of the generalized Langevin Eq.~\eqref{GLEt} reads
\begin{eqnarray*}
&&-i \omega M\vec V(\omega) = -\zeta^\text{\tiny{+}}(\omega) \vec V(\omega) + \vec \xi(\omega)+  \vec{F}_{\text{e}}(\omega)\,,
\end{eqnarray*}
and may be rewritten as
\begin{equation}\label{newton3_bis}
-i \omega M \vec V(\omega) = \vec f (\omega) + \vec{\tilde f}(\omega)+ {\vec F}_{\text{e}}(\omega),
\end{equation}
where we have divided the force exerted by the fluid into deterministic $\vec f(\omega)$ and random $\vec{\tilde f}(\omega)$ components:
\begin{subequations}
\begin{eqnarray}
\vec f(\omega) &\equiv &- \zeta^\text{\tiny{+}}(\omega)  \langle \vec V(\omega)\rangle, \label{hforce1}\\
\vec {\tilde f}(\omega) &\equiv &- \zeta^\text{\tiny{+}}(\omega)  \vec {\tilde V}(\omega) + \vec{ \xi}(\omega) \label{hforce2}\ ,
\end{eqnarray}
\end{subequations}
with $\vec V\equiv \langle \vec V \rangle + \tilde{\vec V}$.  It is easy
to see that $\vec f(\omega)$ is the force exerted by the deterministic
flow field $ \vec u \equiv \langle \vec v \rangle$, the solution of
\begin{subequations}\label{sys_bis}
\begin{eqnarray}
&&i \omega \varrho  \vec u(\vec r, \omega) + \nabla \cdot  \tensor \sigma (\vec r, \omega) = 0\,,\label{sys_bis1}\\
&&\nabla \cdot \vec u (\vec r, \omega) = 0\,, \label{sys_bis2}\\
&&\vec u(\vec r, \omega) = \langle \vec V(\omega)\rangle  \mbox{ on } \mathcal{S} \label{sys_bis3}
\end{eqnarray}
\end{subequations}
while $\tilde{\vec f}(\omega)$  is the force exerted by the stochastic
flow field $\tilde{ \vec u}\equiv  \vec v-\vec \langle \vec v \rangle$, the
solution of  
\begin{subequations}\label{rand_bis}
\begin{eqnarray}
&&i \omega \varrho \tilde{\vec u}(\vec r, \omega) + \nabla \cdot \tilde{\tensor \sigma} (\vec r, \omega) = -\nabla\cdot \tensor \tau(\vec r, \omega)\, ,\label{rand_bis1}\\
&&\nabla \cdot \tilde{\vec u} (\vec r, \omega) = 0\,, \label{rand_bis2}\\
&&\tilde {\vec u}(\vec r, \omega) = \tilde {\vec V}(\omega) \mbox{ on } \mathcal{S}. \label{rand_bis3}
\end{eqnarray}
\end{subequations}
This splitting of equations and boundary conditions is again allowed
by the linearity of the problem. In the following calculation, in
order to ease the notation, we omit the arguments $\vec r$ and
$\omega$ of the hydrodynamic fields where there is no possibility of
confusion.  

We start by calculating twice the energy dissipated by the particle
moving at velocity $\langle \vec V(\omega) \rangle$:
\begin{eqnarray}\nonumber
&&\langle V_i(\omega)\rangle (\zeta^\text{\tiny{+}}(\omega)+{\zeta^\text{\tiny{+}}}^*(\omega)) \langle V_i^*(\omega) \rangle  =\nonumber \\
&\stackrel{\eqref{hforce1}}{=}&-(f_i(\omega)\langle V_i^*(\omega)\rangle+ f_i^*(\omega)\langle V_i(\omega)\rangle)\nonumber \\
&\stackrel{\eqref{formom}}{=}& \langle V_i^*(\omega) \rangle  \int_{\mathcal{S}} \sigma_{ij} n_j \, \diff^2 r +\langle V_i(\omega) \rangle  \int_{\mathcal{S}} \sigma_{ij}^* n_j \, \diff^2 r \nonumber\\
&\stackrel{\eqref{sys_bis3}}{=}&\int_{\mathcal{S}} u_i^* \sigma_{ij} n_j \, \diff^2 r +\int_{\mathcal{S}} u_i\sigma_{ij}^* n_j \, \diff^2 r \label{div} \\
&=&\int_{\mathcal{V}} \partial_j\left(u_i^*\sigma_{ij}\right) \, \diff^3 r \label{div_th}+\int_{\mathcal{V}} \partial_j\left(u_i \sigma_{ij}^*\right) \, \diff^3 r \nonumber \\
&\stackrel{\eqref{sys_bis1}}{=}& \int_{\mathcal{V}} \left( \sigma_{ij} \partial_j u_i^*+ \sigma_{ij}^* \partial_j u_i\right)  \diff^3 r \nonumber\\
&\stackrel{\eqref{sys_bis2}}{=}& \int_{\mathcal{V}}\! 2 \eta \left( \Gamma_{ij} \partial_j u_i^*+ \Gamma_{ij}^* \partial_j u_i\right) \diff^3 r \nonumber\\
&=& 2 \int_{\mathcal{V}} \phi(\vec r, \omega) \, \diff^3 r \label{friction},
\end{eqnarray}
where in Eq.~\eqref{div} we employed the divergence theorem and in Eq.~\eqref{friction} we defined the dissipation function:
\begin{equation*}
\phi(\vec r, \omega)\!\equiv \!  \eta(\vec r) \left(\partial_i u_j \partial_i u_j^* + \partial_i u_j\partial_j u_i^*\right)\!(\vec r, \omega).
\end{equation*}
Since $\zeta(\omega)=2\Re e \zeta^\text{\tiny{+}}(\omega)=\zeta^\text{\tiny{+}}(\omega)+{\zeta^\text{\tiny{+}}}^*(\omega)$, being $\zeta(t)$ real and time-symmetric, we can rewrite Eq.~\eqref{friction} in the following form:
\begin{equation}\label{dissip}
\zeta(\omega)\delta_{ij}\langle V_i(\omega)\rangle \langle V_j^*(\omega)\rangle=2 \int_{\mathcal{V}} \phi(\vec r, \omega) \, \diff^3 r\,
\end{equation}
This proves Eq.~\eqref{result1}. We proceed with the evaluation
of the energy supplied by the random force $\vec \xi(\omega)$:
\begin{eqnarray}
&& \xi_i(\omega) \langle V_i(\omega) \rangle \stackrel[\hphantom{Eq.~\eqref{hforce2}}]{\eqref{hforce2}}{=} \left(\tilde f_i(\omega)+\zeta^\text{\tiny{+}}(\omega) \tilde V_i(\omega)   \right) \langle V_i(\omega) \rangle \nonumber\\
&\stackrel{\eqref{hforce2}}{=}& \tilde f_i(\omega)\langle V_i(\omega) \rangle- f_i(\omega) \tilde V_i(\omega) \nonumber \\
&\stackrel{\eqref{formom}}{=}& -\langle V_i(\omega)\rangle  \int_{\mathcal{S}} \tilde \sigma_{ij} n_j \, \diff^2 r +\tilde V_i(\omega)  \int_{\mathcal{S}} \sigma_{ij} n_j \, \diff^2 r \nonumber\\
&\stackrel{\eqref{sys_bis3}\eqref{rand_bis3} }{=}&-\int_{\mathcal{S}} u_i \tilde \sigma_{ij} n_j \, \diff^2 r +\int_{\mathcal{S}} \tilde u_i\sigma_{ij} n_j \, \diff^2 r \label{div2} \\
&=& -\int_{\mathcal{V}} \partial_j\left(  u_i \tilde \sigma_{ij}\right)\, \diff^3 r  +\int_{\mathcal{V}} \partial_j\left(  \tilde u_i \sigma_{ij}\right)\, \diff^3 r   \nonumber \\
&\stackrel{\eqref{sys_bis1} \eqref{rand_bis1}}{=}& -\int_{\mathcal{V}} \tilde \sigma_{ij} \partial_j  u_i \diff^3 r +\int_{\mathcal{V}} \sigma_{ij} \partial_j \tilde u_i \diff^3 r + \int_{\mathcal{V}} u_i \partial_j \tau_{ij}  \diff^3 r \nonumber \\
&\stackrel{\eqref{sys_bis2} \eqref{rand_bis2}}{=}& \int_{\mathcal{V}} u_i \partial_j \tau_{ij}  \diff^3 r =-\int_{\mathcal{V}} \tau_{ij}  \partial_j u_i \diff^3 r \nonumber.
\end{eqnarray}
In Eq.~\eqref{div2} we made use again of the divergence theorem. Summing up,
\begin{equation}\label{noise}
\xi _i(\omega) \langle V_i(\omega) \rangle = -\int_{\mathcal{V}}  \tau_{ij}(\vec r, \omega)  \partial_j u_i(\vec r,\omega) \diff^3 r,
\end{equation}
which asserts that $\vec \xi (\omega)$ is Gaussian with vanishing
mean, being the integral of the deterministic quantity $\partial_j
u_i$ times the zero-mean Gaussian field $ \tau_{ij}$. 
Using Eq.~\eqref{noise} we evaluate the noise correlation function : 
\begin{eqnarray}
&&\langle\xi_i(\omega)  \xi_j^{*}(\omega')\rangle \langle V_i(\omega)\rangle \langle V_j^*(\omega')\rangle=\nonumber \\
&=& \int_{\mathcal{V}} \diff^3 r' \int_{\mathcal{V}} \diff^3 r  \, \partial_j u_i(\vec r,\omega)  \langle \tau_{ij}(\vec r, \omega) \tau_{kl}^*(\vec r', \omega') \rangle \partial_l u_k^*(\vec r',\omega') \nonumber\\
&=& 2 \kB \delta(\omega-\omega') \int_{\mathcal{V}}  \eta \left(\partial_i u_j \partial_i u_j^*+ \partial_i u_j\partial_j u_i^*\right) T \, \diff^3 r \label{noiseft}\\
&=& 2 \kB \delta(\omega-\omega') \int_{\mathcal{V}}  \phi(\vec r, \omega) T(\vec r) \, \diff^3 r \nonumber.
\end{eqnarray}
In Eq.~\eqref{noiseft} we used the Fourier transform of Eq.~\eqref{fluidfdt} together with $\tau_{kl}^*(\vec r', \omega')=\tau_{kl}(\vec r', -\omega')$, since $\tensor \tau$ is real. This proves Eq.~\eqref{result2}.

\section{Hydrodynamics of a translating and rotating sphere}\label{sec:hydro}

\subsection{Translational motion}
The Fourier transform of the flow field generated by a sphere translating with velocity $V(\omega)\vec e_{z}$ reads  in polar coordinates $(r, \varphi,\theta)$  \cite[p.~623]{Reichl.1998}:
\begin{equation*}
\vec u^{\text{\tiny{T}}}(r, \theta,\omega)= \frac{1}{r}\left[ \sin\theta\left(g+ r\frac{\diff g}{\diff r}\right)\!\vec e_\theta-2g\cos\theta  \vec e_r\right], \label{transl}
\end{equation*}
with
\begin{equation*}
g(r,\omega)= \frac{3\eta V(\omega)R}{2(kr)^2} \left[ \left(ik-1\right)e^{ik(r-R)}-\!\left(1+ikR-\frac{1}{3}(kR)^2\right) \right],
\end{equation*}
where $k = (1 + i) k_0$, and $k_0 = \sqrt{\omega/2 \nu}$ is the inverse of the characteristic fluid diffusion length.
The associated dissipation function is:
\begin{equation*}
\phi^{\text{\tiny{T}}} = \eta \left( \frac{12}{r^4}\cos^2\theta\left| g- r\frac{\diff g}{\diff r}  \right|^2+ \sin^2\theta\left|\frac{\diff^2 g}{\diff r^2}\right|^2\right)\,,  
\end{equation*}
that becomes after integration over $\theta$:
\begin{eqnarray}
&&\int_{0}^{\pi} \phi^{\text{\tiny{T}}}(r,\theta,\omega) \sin\theta \diff \theta= \frac{3\eta |V(\omega)|^2 R^2}{2 k_0^4 r^8} \times \nonumber \\ 
&& \left\lbrace 5 [9+2 k_0 R (9+k_0 R (9+2 k_0 R (3+k_0 R)))]+e^{-2 k_0 (r-R)} [45+2 k_0 r (45+k_0 r (45+k_0 r (30+k_0 r (15+2 k_0 r (3+k_0 r)))))] \right.\nonumber \\
&& \left.-2 e^{-k_0 (r-R)} \left[\left(45+45 k_0 R+15 k_0 (3+2 k_0 R (3+k_0 R)) r+12 k_0^3 R (3+2 k_0 R) r^2+2 k_0^3 \left(-3+2 k_0^2 R^2\right) r^3\right) \cos[k_0 (R-r)] \right. \right. \nonumber \\
&&  \left. \left. -k_0 \left(-15 R (3+2 k_0 R)+15 \left(3-2 k_0^2 R^2\right) r+36 k_0 (1+k_0 R) r^2+2 k_0^2 (3+2 k_0 R (3+k_0 R)) r^3\right) \sin[k_0 (R-r)]\right] \vphantom{e^{-k_0(R-r)}}\right\rbrace. \label{phi_transl}
\end{eqnarray}
Notice that Eq.~\eqref{phi_transl} displays a term which does not decay
with an exponential cutoff but only algebraically as $1/r^8$.  But its
contribution to $\mathcal{T}(\omega)$ actually diminishes at high
frequencies, $k_0 \to \infty$. In order to obtain
$\mathcal{T}(\omega)$ we numerically integrate
Eq.~\eqref{Tdef} together with Eq.~\eqref{phi_transl}. The result is shown in
Figure \ref{fig:Tomega}.

\subsection{Rotational motion}
The Fourier transform of the flow field generated by a sphere rotating  with angular velocity $\Omega(\omega)\vec e_{z}$  reads \cite[p.~91]{Landau.1987}:
\begin{equation}\nonumber
\vec u^{\text{\tiny{R}}}(r, \theta,\omega)= \frac{\Omega(\omega) R^3}{r^2} \sin \vartheta \frac{1 - i k r}{1 - i k R} e^{i [k (r - R) ]} \, \vec e_\varphi \equiv  f(r,\vartheta,\omega) \vec e_\varphi.\label{rot}
\end{equation}
The associated dissipation function $\phi^{\text{\tiny{R}}}(\vec r, \omega)$ is:
\begin{align}
\phi^{\text{\tiny{R}}} &= \frac{\eta}{r^2} \left( \left|r \partial_r
    f - f\right|^2 + \left|\partial_\vartheta f - \cot \vartheta
    f\right|^2\right) \nonumber \\
    &=  \frac{\eta |\Omega(\omega)|^2 R^6 \left[9 + 18
    k_0 r + 18 (k_0 r)^2 + 12 (k_0 r)^3 + 4 (k_0 r)^4\right]
      \sin^2 \!\vartheta e^{- 2 k_0 (r - R)}}{r^6
  \left[1 + 2 k_0 R + 2 (k_0 R)^2   \right]}\,. \label{phi_rot}
\end{align}
Notice in Eq.~\eqref{phi_rot} the exponential cutoff where the fluid's
diffusion characteristic length $k_0^{-1}$ appears.  Using
Eq.~\eqref{Tdef} and Eq.~\eqref{phi_rot} we obtain the first-order
approximation in $\Delta T$ for the noise temperature of
rotational motion:
\begin{equation}
\frac{\mathcal{T}^{\text{\tiny R}}(\omega) - T_0}{\Delta T} = \frac{9
  + 18 k_0 R + 18 (k_0 R)^2 + 12 (k_0 R)^3 - 8 (k_0 R)^4 E_1(2 k_0
  R)e^{2 k_0 R} }{ 4\left(3 + 6 k_0 R + 6 (k_0 R)^2 + 2 (k_0
    R)^3\right)}\,,\label{deltaT}
\end{equation}
where $E_1(x) = \int_1^\infty \diff y\, e^{- x y}/y$ is the exponential integral. The result is plotted in Figure \ref{fig:Tomega}.
\end{widetext}

\bibliography{LongHBM} 
\bibliographystyle{apsrev4-1}

\end{document}